# The Green Bank North Celestial Cap Survey. IX. Timing Follow-up for 128 Pulsars

A. E. McEwen[1,2], J. K. Swiggum[1,3], D. L. Kaplan[1], C. M. Tan[4,5], B. W. Meyers[6,7], E. Fonseca[8,9], G. Y. Agazie[1], P. Chawla[4], K. Crowter[7], M. E. DeCesar[10,27], T. Dolch[11,12], F. A. Dong[7], W. Fiore[13,14], E. Fonseca[13,14], D. C. Good[15], A. G. Istrate[16], V. M. Kaspi[4,5], V. I. Kondratiev[17], J. van Leeuwen[17], L. Levin[18], E. F. Lewis[13,14], R. S. Lynch[19], K. W. Masui[20,21], J. W. McKee[22,23], M. A. McLaughlin[13,14], H. Al Noori[24], E. Parent[4,5], S. M. Ransom[25], X. Siemens[26], R. Spiewak[18], and I. H. Stairs[7]

[1] Center for Gravitation, Cosmology, and Astrophysics, Department of Physics, University of Wisconsin-Milwaukee, P.O. Box 413, Milwaukee, WI 53201, USA
[2] Department of Physics, The George Washington University, Washington, DC 20052, USA
[3] Department of Physics, 730 High Street, Lafayette College, Easton, PA 18042, USA
[4] Department of Physics, McGill University, 3600 University Street, Montréal, QC, H3A 2T8, Canada
[5] McGill Space Institute, McGill University, 3550 University Street, Montréal, QC, H3A 2A7, Canada
[6] International Centre for Radio Astronomy Research (ICRAR), Curtin University, Bentley WA 6102 Australia
[7] Department of Physics and Astronomy, University of British Columbia, 6224 Agricultural Road, Vancouver, BC, V6T 1Z1 Canada
[8] Department of Physics and Astronomy, West Virginia University, P.O. Box 6315, Morgantown, WV 26506, USA
[9] Center for Gravitational Waves and Cosmology, West Virginia University, Chestnut Ridge Research Building, Morgantown, WV 26505, USA
[10] George Mason University, Fairfax, VA 22030, USA
[11] Department of Physics, Hillsdale College, 33 E. College Street, Hillsdale, MI 49242, USA
[12] Eureka Scientific, 2452 Delmer Street, Suite 100, Oakland, CA 94602-3017, USA
[13] Department of Physics and Astronomy, West Virginia University, Morgantown, WV 26506, USA
[14] Center for Gravitational Waves and Cosmology, West Virginia University, Chestnut Ridge Research Building, Morgantown, WV 26506, USA
[15] Department of Physics and Astronomy, University of Montana, 32 Campus Drive, Missoula, MT 59812 USA
[16] Department of Astrophysics/IMAPP, Radboud University, P.O. Box 9010, 6500 GL Nijmegen, The Netherlands
[17] ASTRON, the Netherlands Institute for Radio Astronomy, Oude Hoogeveensedijk 4, 7991 PD Dwingeloo, The Netherlands
[18] Jodrell Bank Centre for Astrophysics, School of Physics and Astronomy, The University of Manchester, Manchester, M13 9PL, UK
[19] Green Bank Observatory, P.O. Box 2, Green Bank, WV 24494, USA
[20] MIT Kavli Institute for Astrophysics and Space Research, Massachusetts Institute of Technology, 77 Massachusetts Avenue, Cambridge, MA 02139, USA
[21] Department of Physics, Massachusetts Institute of Technology, 77 Massachusetts Ave, Cambridge, MA 02139, USA
[22] E.A. Milne Centre for Astrophysics, University of Hull, Hull, HU6 7RX, UK
[23] Centre of Excellence for Data Science, Artificial Intelligence and Modeling, University of Hull, Hull, HU6 7RX, UK
[24] Department of Physics, University of California, Santa Barbara, CA 93106, USA
[25] National Radio Astronomy Observatory, 520 Edgemont Road, Charlottesville, VA 22903, USA
[26] Department of Physics, Oregon State University, Corvallis, OR 97331, USA

## Abstract

The Green Bank North Celestial Cap survey is one of the largest and most sensitive searches for pulsars and transient radio objects. Observations for the survey have finished; priorities have shifted toward long-term monitoring of its discoveries. In this study, we have developed a pipeline to handle large data sets of archival observations and connect them to recent, high-cadence observations taken using the Canadian Hydrogen Intensity Mapping Experiment telescope. This pipeline handles data for 128 pulsars and has produced measurements of spin, positional, and orbital parameters that connect data over observation gaps as large as 2000 days. We have also measured glitches in the timing residuals for five of the pulsars included and proper motion for 19 sources (13 new). We include updates to orbital parameters for 19 pulsars, including nine previously unpublished binaries. For two of these binaries, we provide updated measurements of post-Keplerian binary parameters, which result in much more precise estimates of the total masses of both systems. For PSR J0509+3801, the much improved measurement of the Einstein delay yields much improved mass measurements for the pulsar and its companion, 1.399(6) $M_\odot$ and 1.412(6) $M_\odot$, respectively. For this system, we have also obtained a measurement of the orbital decay due to the emission of gravitational waves, $\dot{P}_{\rm B} = -1.37(7) \times 10^{-12}$, which is in agreement with the rate predicted by general relativity for these masses.

*Unified Astronomy Thesaurus concepts:* Radio pulsars (1353); Binary pulsars (153)

## 1. Introduction

Since the discovery of pulsars in the late 1960s (Hewish et al. 1968), they have been the subjects of intense study. As these stars rotate, their magnetic fields beam radio waves from their poles like a lighthouse. On Earth, this rotation is detectable as a series of pulses. Precise measurements of these pulses at Earth track the passage of time in the pulsar's frame, and the timing models compare this clock to those on Earth. Deviations from the predicted arrival times of pulses encode information about the pulsar and material along the line of sight to it in a process called "pulsar timing" (e.g., Lorimer & Kramer 2012).

The astrophysical applicability of pulsars is wide (e.g., Backer 1975, 1984; Hewish 1975; Detweiler 1979; Taylor & Weisberg 1982; Taylor et al. 1992). For instance, pulsar astronomy has provided some of the most stringent constraints on the behavior of ultra-dense matter (Özel & Freire 2016; Bogdanov et al. 2019; Hu et al. 2020; Lattimer 2021; Pang et al. 2021). Their immense moments of inertia make them some of the most stable clocks in the Universe (Manchester 2017; Yin et al. 2017). The most stable millisecond pulsars (MSPs) can be used as an ensemble to search for low-frequency gravitational waves (Detweiler 1979; Desvignes et al. 2016; Reardon et al. 2016; Arzoumanian et al. 2020; Antoniadis et al. 2022; Agazie et al. 2023a, 2023b, 2023c, 2023d, 2023e; Afzal et al. 2023).

---

[27] Resident at the US Naval Research Laboratory, Washington, DC 20375, USA.

All these topics benefit from ongoing searches for pulsars, which continue to increase the known population. These searches implement ever-improving technology and searching algorithms that have discovered over 3000 pulsars to date (e.g., Manchester et al. 1978, 2001, 2005; Deneva et al. 2013; Stovall et al. 2014; Sanidas et al. 2019; Cruces et al. 2021; Sengar et al. 2023, etc.). Many of these pulsars have been timed to high precision, and have been used to characterize the pulsar population (Faucher-Giguère & Kaspi 2006; Lorimer et al. 2006; Levin et al. 2013; Lorimer et al. 2015).

This timing process is hampered by many components of pulsar evolution. Young pulsars are born following the supernova of massive progenitors; these highly magnetized neutron stars rotate the least stably and slow the most quickly (Antonelli et al. 2023). This slowing is due at least in part to magnetic dipole radiation, but typically does not follow so simple a model. Many pulsars, especially young pulsars, exhibit some form of timing noise (Parthasarathy et al. 2019; Lower et al. 2020; Singha et al. 2022). This can manifest as a long-term, random drift in the observed spin of a pulsar that deviates from "pure" spin-down due to magnetic dipole radiation (a power law in spin with an index of 3). Measurements have shown that this model is very rarely sufficient, suggesting a more complex relationship between these parameters (Lower et al. 2020). More generally, the adolescent pulsar's spin will follow a random walk in $\nu$ and $\dot\nu$. From the timing side, this noise is often removed by fitting and subtracting polynomials from the residuals—a process known as "polynomial whitening" (Hobbs et al. 2004).

Along with this drift, some pulsars show abrupt changes in spin period called "glitches" (Espinoza et al. 2011; Fuentes et al. 2017; Zhou et al. 2022a, 2022b; Basu et al. 2022). Theories to explain these events largely utilize either of the following:

1. A starquake model where the crust of the neutron star abruptly changes shape and the change in its moment of inertia drives the observed momentum-conserving change in rotation (Alpar et al. 1994; Lai et al. 2018; Lu et al. 2023).
2. Differential rotation of the solid crust and a superfluid interior that is coupled by the pinning of vortices in this superfluid (Alpar 1977; Link et al. 1992; Haskell et al. 2020; Layek et al. 2023; Melatos & Millhouse 2023).

While the process that causes pulsars to glitch has still not yet been definitively identified, their impacts on timing have been addressed in many studies (e.g., Lower et al. 2021; Dunn et al. 2022; Zubieta et al. 2023). In a simple case, the glitch appears in timing residuals as an instantaneous change in the slope. This change is also accompanied by a change in the spin-down rate of the pulsar, which introduces an additional accumulation of phase offset in the residuals. Many observed glitches are then followed by an exponential recovery of the affected parameters back to their pre-glitch values. All of these components manifest as changes in pulsar spin parameters, and complicate measurements of spin/spin-down.

On the other hand, the rotation of old pulsars is typically much more stable, particularly those that reside in binary systems. In many cases, these pulsars will harvest rotational momentum from the orbit via the Roche-lobe overflow of the companion. This process decreases the rotation period of the pulsar, potentially down to the millisecond level. It also greatly dampens their magnetic fields, reducing the long-term spin-down rate from dipole radiation (Bisnovatyi-Kogan & Komberg 1974; Shibazaki et al. 1989). These pulsars, called millisecond pulsars (or MSPs), have been utilized for the majority of high-precision pulsar science (Alam et al. 2021; Arzoumanian et al. 2023; Falxa et al. 2023; Miles et al. 2023). Aside from their stability, the timing procedure is sensitive to Doppler shifts due to binary motion. These shifts can be used to determine three of the Keplerian parameters of the orbit, including the orbital period, the pulsar's semimajor axis, and the initial phase of the binary (relative to the time of measurement). These can be used to place limits on the masses in the system via the Keplerian mass function. Some binary orbits are compact enough that the pulsar reaches relativistic speeds during the orbit, introducing additional complexities in the timing procedure that can be used to directly measure the mass of the pulsar and its companion (Demorest et al. 2010; Özel & Freire 2016; Cromartie et al. 2020). Given that this measurement is nearly impossible to make otherwise, binary pulsars are unique and powerful laboratories for gravitation and ultra-dense matter; and in those few sources where the masses can be measured via detection of multiple independent post-Keplerian parameters (as for the double pulsar system PSR J0737−3039 and the triple system PSR J0337+1715), stringent constraints are placed on the underlying theory (Archibald et al. 2018; Voisin et al. 2020; Kramer et al. 2021).

Precise measurements of these parameters is hampered by infrequent sampling, as models that describe pulse arrival times well over a single observation may fail to do so for subsequent observations when there are long wait times between scans. The development of high-cadence radio instruments like the Canadian Hydrogen Intensity Mapping Experiment (CHIME; CHIME/Pulsar Collaboration et al. 2021), which observe large swathes of the sky on a daily basis, helps to avoid this loss of phase connection. In this way, CHIME has already proven itself a very capable pulsar-timing instrument, especially with sources that are difficult to time with episodic timing campaigns (Fonseca et al. 2021; Good et al. 2021; Dong et al. 2023).

Here, we discuss the continued efforts of a pulsar survey that has reached the end of its observing program: the Green Bank North Celestial Cap pulsar survey (GBNCC). We discuss the survey's conclusion and current objectives in Section 1.1. In Section 2, we describe the sample of pulsars we have examined in this study and how we produce data products, including the addition of high-cadence CHIME observations. Section 3 outlines the procedure used to generate timing data products and use them to refine models. Section 4 covers the results from our study and highlights some particularly interesting measurements. Finally, Section 5 summarizes our results. We also include an Appendix with timing residuals for all sources. The work below includes comparisons with the ATNF Pulsar Catalog (v1.70; Manchester et al. 2005).[28]

### 1.1. GBNCC Survey Overview and Completion

Over the last decade, the GBNCC survey has covered the sky north of $\delta = -40°$ with 124,852 pointings of 2 minutes duration with the 100 m Green Bank Telescope (GBT) at 350 MHz. Full specifications of survey observations and processing are detailed in previous GBNCC publications (Stovall et al. 2014; Lynch et al. 2018b). This choice of central frequency provides additional sensitivity to steep spectrum sources, in particular older pulsars outside of the Galactic plane (McEwen et al. 2020). The survey is one of the largest pulsar surveys in terms of sky coverage, covering 36,430 deg$^2$.

To date, the GBNCC survey has discovered 195 sources, including 33 MSPs, 24 binaries, 24 rotating radio transients (McLaughlin et al. 2006), and a fast radio burst (FRB; Lorimer 2018), many of which have been published in a series of survey papers (Stovall et al. 2014; Lynch et al. 2018b; Kawash et al. 2018; Aloisi et al. 2019; Parent et al. 2020; Agazie et al. 2021; Fiore et al. 2023; Swiggum et al. 2023). These discoveries have included some of pulsar astronomy's most exotic sources and important measurements, including the high mass of PSR J0740 +6620 (Cromartie et al. 2020; Fonseca et al. 2021), eclipsing black widow pulsars that are ablating their companions (Swiggum et al. 2023), nulling pulsars (Anumarlapudi et al. 2023), and double neutron star binaries (Lynch et al. 2018b; Aloisi et al. 2019; Swiggum et al. 2023). These studies and others have amassed data on GBNCC discoveries, much of which is currently available from the public-facing GBNCC GitHub repository.[29]

During the summer of 2022, the final GBNCC survey positions were observed (aside from ⩽500 points which will be reobserved for reasons related to radio frequency interference (RFI) and observation complications). Efforts to identify new pulsars are continuing, both with our existing pipeline and with new pipelines that improve sensitivity to certain regions of parameter space (Sengar et al. 2023). For this study, we instead focus on the continued timing of prior discoveries by connecting archival data with new observations. Note that there are also 12 sources included in this paper that were discovered in the GBT350 survey of the northern Galactic plane (Hessels et al. 2008). They were followed up at 820 MHz using the GBT during an earlier timing campaign (project code AGBT13B_290; PI: R. Rosen), and are included among the CHIME/Pulsar sources (for details about the CHIME/Pulsar system, see CHIME/Pulsar Collaboration et al. 2021).

## 2. Sample Assembly

### 2.1. Observations

Following identification, candidates were observed in test scans using either the GBT (earlier discoveries) or CHIME/Pulsar (starting in early 2020). In the former case, two to three 10 minute scans were taken over the course of a few weeks during other survey observations to confirm discoveries. New pulsars were then included in follow-up proposals with the GBT to establish phase-connected timing solutions. The observing specifications of these proposals vary slightly to account for complex timing models, but typically included ≃10 follow-up observations (10–20 minutes on source per scan) with the GBT using either the 350, 820, or 1400 MHz receiver. Depending on the success of the timing campaign and potential scientific benefits, some pulsars were included in subsequent proposals using other instruments like the Low Frequency Array (LOFAR; van Haarlem et al. 2013) and Arecibo.

Following an agreement between the CHIME/Pulsar and GBNCC collaborations in late 2019, candidates (and pulsars) identified in GBNCC data could be followed up using the CHIME/Pulsar instrument with minimal latency due to its continuous and commensal manner of operation. During the spring semester of 2020, many previously discovered (and in some cases, previously published) GBNCC pulsars were observed using the 600 MHz receiver on CHIME with cadences of 1–7 days. The scan length depends on the declination of the source, as pulsars close to the southern horizon are visible for a shorter time compared to those directly overhead. At a minimum (i.e., $\delta \lesssim 0°$), scans were ⩽10 minutes; the longest scans lasted approximately 1 hr. Despite the slight decrease in sensitivity between the GBT versus CHIME, the high cadence of these observations provides unprecedented sensitivity to the pulsar behavior, and models can be updated on a regular basis. The reduction in beam size further constrains the error in position for the sources, particularly when gridding observations were taken. Normal 350 MHz survey observations continued during the intervening years, and new candidates were confirmed/rejected using both GBT test scans and CHIME/Pulsar daily observations. As the CHIME/Pulsar schedule filled, the priority (and cadence) of some sources was reduced to accommodate those with incomplete/exotic models. Eight pulsars discovered with declinations below CHIME's horizon ($-10°$) were also included in our timing analysis. These sources were observed solely with the GBT in dedicated timing campaigns.

### 2.2. GBNCC/CHIME Pipeline

As part of the observing activities automatically undertaken by CHIME/Pulsar, a simplified data-reduction and pulse time-of-arrival (TOA) extraction procedure is applied to each observation in order to assess data quality and generate initial timing data products. The procedure utilizes a Python-based workflow that interacts with the data via the standard PSRCHIVE (van Straten et al. 2012) Python interface. The automated workflow, running on the Cedar supercomputing cluster,[30] will identify any new observation data products that were transferred from the observatory to the storage system. Subsequently, standardized RFI mitigation is performed on the data using clfd (Morello et al. 2019), including the excision of known corrupted frequency channels for that observation obtained via CHIME/Pulsar utilities. Two reduced copies of the cleaned archive are then retained, one with 32 subbands and 1 minute subintegrations, and another one with 1024 frequency channels and fully averaged in time.

Using daily CHIME/Pulsar observations greatly increased the data volume for many GBNCC sources. To best utilize these data, a pipeline was established between the data taking and data processing via GitHub. As new data are taken, single-epoch and frequency-averaged TOAs are generated using preliminary timing models and a detection pulse template. These TOAs are then uploaded in batches to a shared private

---

[28] http://www.atnf.csiro.au/research/pulsar/psrcat.
[29] https://github.com/GBNCC/data

[30] Operated by the Digital Research Alliance of Canada, https://docs.alliancecan.ca/wiki/Cedar.

GitHub repository, where they can be accessed by GBNCC. With the new data, timing models are updated using fitting tools in PINT (Luo et al. 2021). When a solution has been updated, it can then be passed back to CHIME/Pulsar for future observations as necessary. In some cases, new timing models change the pulse profile enough to justify updating the TOA template, and therefore requires re-extraction of TOAs from all previous observations. Phase-connected solutions offer positional precision well within the observing beam size ($\sim 0.5°$ at 400 MHz, $\sim 0.25°$ at 800 MHz), particularly after a year of observing.

## 3. Data Analysis

### 3.1. TOA Excision

The data set contains 78,585 TOAs from four telescopes (GBT, Arecibo, LOFAR, and CHIME). As the timing models develop, measured TOA uncertainties will reduce as the profile is improved. In some cases, position improvements from phase connection dramatically improve the signal-to-noise ratio (S/N) of detections. In others, individual CHIME observations resulted in low-S/N detections, even with a timing-derived position. In these cases, multiple observations were summed together before producing TOAs, resulting in fewer (but more precise) TOAs. The wealth of data coming from CHIME allows us to be more selective with TOAs, and we implement TOA zapping as a part of the timing pipeline. Many of the zapped TOAs are due to faint detections and excess RFI. There are also a few epochs of known TOA issues which have been removed more broadly from our data set. As discussed in Andersen et al. (2023), improper packaging of CHIME/Pulsar data prior to MJD 58550 resulted in corrupted data. These TOAs were zapped from all of the sources.

### 3.2. Timing Procedure

The timing baselines for pulsars included in this data set are highly source dependent. A large number of sources (95) include archival data (pre-2019) from the GBT and other telescopes, and nearly all (120) sources have data taken with CHIME (post-2019). There are eight pulsars not visible with CHIME, and so solutions were determined from GBT observations only. For this project, data are only included up to 2023 May, and a minimum baseline of 1 yr was imposed.

After excising corrupted data, predicted pulse TOAs were compared with the time stamps of observed pulses. Correlated deviations indicate errors in the timing model that are fit using a least-squares algorithm. Determining the appropriate parameters to include in this fit is difficult, as covariance is high for any parameters that influence the residuals on timescales larger than the baseline. The covariance of sinusoidal (position, velocity, and orbital motion) and quadratic (rotation rate and deceleration) components limits the prediction power of timing models when the baseline is short. In many cases, the gap between GBT timing campaigns and CHIME/Pulsar daily scans is >2000 days. Fitting models over this gap can be very constraining, but requires a long enough baseline on one side of the gap to reduce phase uncertainty at the other side to within a single rotation.

When starting a timing solution from scratch, a preliminary solution with discovery parameters is used to fit TOAs over a short time period (typically a single observation). This fit only constrains the rotation rate of the pulsar as it is the only parameter with significant influence on short timescales. The improved frequency is then used to model TOAs at the MJD of the nearest observation, and a fit to both days provides a more precise measurement. The regular cadence of CHIME/Pulsar observations can introduce a subtle aliasing error into the model, as the pulsar rotation can become degenerate with the observation spacing determined by sidereal rate (see in CHIME/FRB Collaboration et al 2020). To mitigate this, single CHIME observations can be split into multiple TOAs to find an appropriate starting model to fit to subsequent days. This procedure is repeated until the model no longer appropriately fits the data and additional parameters must be included. When multiband data exist for a pulsar, additional constraints are set on its dispersion measure (DM). This constraint benefits from CHIME's wide observing band. Errors in DM lead to a smeared pulse and reduced S/N, so implementing new DM measurements further improves TOA precision.

Included among the pulsars in this project are 19 binary pulsars. These sources require special attention in the timing process, especially when the orbital period is $\geqslant 1$ yr. For short orbital periods (where the observation duration $t_{\rm obs}$ spans $\geqslant 1$ orbital period $P_{\rm B}$), orbital parameters can be constrained almost immediately afterwards. When the orbital period is long ($P_{\rm B} \geqslant$ 1 yr), covariance with Earth's motion and the pulsar's intrinsic spin-down become significant. For these cases, fitting timing residuals with polynomial functions of order $\geqslant 2$ captures the apparent spin derivatives due to the orbital motion, and can be used to approximate the orbital parameters. Ultimately, these derivatives are replaced by orbital parameters when possible.

## 4. Results

We have assembled timing data for 128 sources; timing residuals for all sources are given in the Appendix and a period–period derivative plot is shown in Figure 1. Many of the sources included have had detailed timing analyses published in previous/upcoming GBNCC publications. For these sources, we focus on maintaining phase coherence over long baselines and measurements of new parameters. For others, we provide new timing-derived solutions for spin and position parameters.

### 4.1. New Pulsars

The majority of pulsars timed in this study have been published in preceding GBNCC survey papers (Stovall et al. 2014; Lynch et al. 2018a; Kawash et al. 2018; Aloisi et al. 2019; Agazie et al. 2021; Swiggum et al. 2023), but there are 30 sources that have been discovered within the past 2–3 yr for which no timing solution has been previously available. These sources, listed in Tables 1 and 2, include a wide range of periods and DMs. Because they were discovered following the CHIME/GBNCC data sharing agreement, many of these sources only have TOAs from CHIME observations. Along with the GBNCC discoveries, we provide timing solutions for the sources that were discovered in the GBT350 Northern Galactic Plane (NGP) survey (Hessels et al. 2008). These 12 sources are marked in Tables 1 and 2.

### 4.2. Glitches and Timing Noise

In this data set, sources with spin derivatives of order >2 and/or apparent discontinuities in timing residuals were examined as possible glitching pulsars. To model such a glitch, up to four parameters are used: the epoch of the glitch (given in MJD), the magnitude of the change in spin frequency ($\Delta f$, given in hertz), and the magnitude of the change in first and second frequency derivatives ($\Delta \dot{\nu}$, in Hz s$^{-1}$ and $\Delta \ddot{\nu}$ in Hz s$^{-2}$). In some cases (especially those where the glitch occurs in a low-observing-cadence epoch), only one of these parameters can be reliably measured.

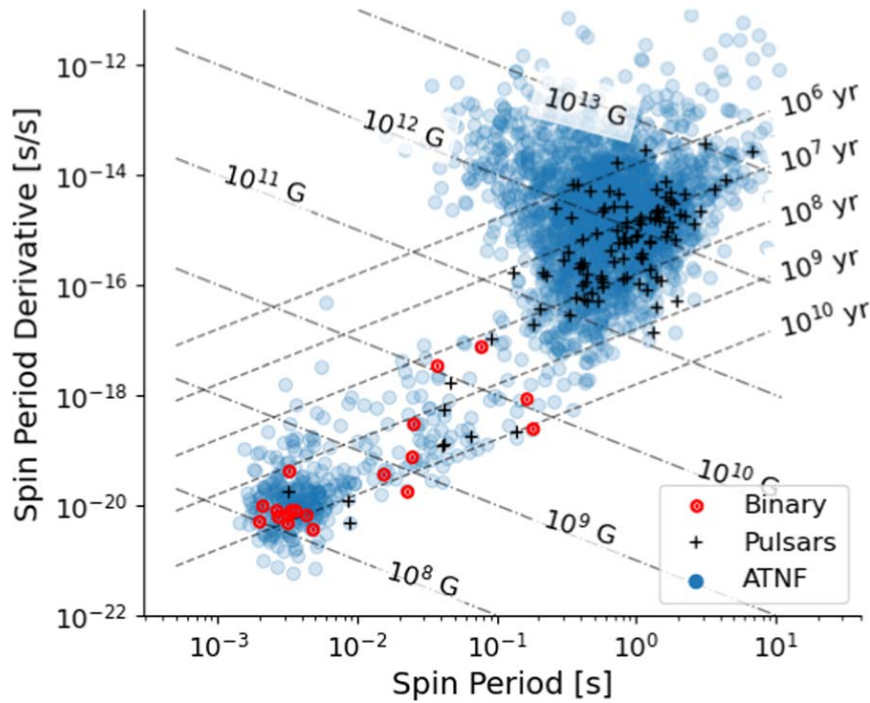

**Figure 1.** Spin period vs. period derivative for all sources timed in this work. Black and red markers indicate sources we have timed, with the latter indicating binary sources. We also plot all pulsars from the ATNF Pulsar Catalog (Manchester et al. 2005). Dashed and dashed–dotted lines indicate constant characteristic ages and magnetic fields, respectively.

With a high observing cadence (and a high-significance glitch with magnitude typically of order $\geqslant 10^{-10}\nu$), glitches are relatively simple to resolve: A precise enough pre-glitch model can make the sudden period change quite stark. Sources that experienced glitches during the CHIME/Pulsar data are in this regime, as near-daily cadences greatly reduce our uncertainty on the glitch epoch. However, many of the sources included in this project have two or more temporally disparate observing epochs, and observations on either side of gaps are generally taken using different instruments. Distinct observing systems can introduce additional pulse phase uncertainty due to both instrumental and profile evolution effects; to mitigate this, we typically include a free "jump" parameter between different systems/frequencies. This greatly improves fits between multiple telescopes, but it does reduce sensitivity to glitches in the gaps, as the jump is highly covariant with the glitch epoch. We also acknowledge that the GBNCC survey has been shown to be more sensitive to older pulsars (McEwen et al. 2020), which are less likely to glitch than their younger counterparts (Espinoza et al. 2011; Millhouse et al. 2022). Despite this, the long baselines (and large sample of pulsars) increase the odds of observing these phenomena even in unlikely sources.

To that end, we find some pulsars in our data set seem to contain evidence for glitches and report on their characteristics here. We compare the measured glitch parameters to all published glitches in the ATNF glitch database in Figure 2 and find that all measured glitch magnitudes appear well in line with previous measurements, both in $\Delta\nu$ and $\Delta\dot{\nu}$.[31] Fit parameters for all included glitches are given in Table 3, and residuals before and after adding glitches are shown in Figures 3–5.

---

[31] https://www.atnf.csiro.au/people/pulsar/psrcat/glitchTbl.html

### 4.2.1. J1923+4243

PSR J1923+4243 was included in two timing campaigns in 2013 (Lynch et al. 2018b) prior to observations at CHIME. This campaign resulted in a well-constrained timing model, making phase connection to the CHIME/Pulsar observations that began in 2020 trivial. However, following MJD ≃59500, the solution began to drift, and the rms of the residuals began to grow. Initial attempts included adding higher-order frequency derivatives; while this did improve the fit, the parameters were not stable, and changed as more data were added. We substituted the frequency derivative terms for glitch parameters (glitch epoch and glitch magnitude), and the fit dramatically improved (reduced $\chi^2 = 64$, with 333 degrees of freedom, DOF, initially and 2.3 for 331 DOF). We used an F-test to compare the fits; comparing the model after adding glitch parameters to the model without a glitch, the F-statistic is $\approx 10^{-241}$, indicating that the fit improvement is unlikely due to chance.

While this fit is much better than pre-glitch with spin frequency derivatives, some additional structure remained in the residuals. In an attempt to reduce this, we also tried a model that included terms for the second frequency derivative and glitch recovery (first and second frequency derivatives). Given that glitches typically correlate with significant timing noise (Lower et al. 2021), this choice is reasonable. The resulting fit no longer contained any apparent structure, the reduced $\chi^2$ dropped to 1.2 (DOF now 329), and an F-test comparing it to the pre-glitch fit returned $\approx 10^{-281}$. Comparing instead to the simpler glitch model (containing only the epoch and magnitude of the glitch), the F-statistic is $\approx 10^{-44}$ in preference for the more complex model. We plot the timing residuals before and after including the glitch model in Figure 6, and the parameters for the glitch are given in Table 3.

Table 1
Spin Parameters for Newly Solved Pulsars

| PSR | $\nu$ (Hz) | $\dot{\nu}$ ($10^{-14}$ Hz s$^{-1}$) | $\dot{E}$ ($10^{30}$ erg s$^{-1}$) | $\tau_c$ (Myr) | $B_S$ ($10^8$ G) |
|---|---|---|---|---|---|
| J0032+6946 | 27.1711189496937(3) | −0.2650134(3) | 2842.7 | 162.4 | 116.3 |
| J0054+6650[a] | 0.719311913668(6) | −0.287226(5) | 81.6 | 4.0 | 28111.6 |
| J0110−2223 | 0.79241833567(2) | −0.039(2) | 12.2 | 32.3 | 8947.4 |
| J0120+1837 | 0.761391192926(2) | −0.00079(2) | 0.2 | 1533.7 | 1350.8 |
| J0141+6303 | 21.422324384511(12) | −0.075548(12) | 638.9 | 449.3 | 88.7 |
| J0406+3039 | 383.339448073778(9) | −0.122083(8) | 18475.6 | 4975.0 | 1.5 |
| J0415+6111 | 2.27174930502(2) | −0.02587(2) | 23.2 | 139.1 | 1503.3 |
| J0420+4451[a] | 0.805719788443(4) | −0.038462(4) | 12.2 | 33.2 | 8677.4 |
| J0530−3847 | 1.10314158660(2) | −0.0756(2) | 32.9 | 23.1 | 7594.2 |
| J0749+5720 | 0.8511150531782(2) | −0.006017(2) | 2.0 | 224.1 | 3161.2 |
| J0758−3002 | 0.915860185362(9) | −0.0322(2) | 11.7 | 45.0 | 6554.5 |
| J1327+3423 | 24.0890080713045(2) | −0.0075032(2) | 71.4 | 5086.8 | 23.4 |
| J1354+2453 | 1.174999892192(7) | −0.019339(9) | 9.0 | 96.3 | 3493.9 |
| J1602−1009 | 320.9472096060605(97) | −0.04918(4) | 6231.0 | 10340.4 | 1.2 |
| J1604−0057 | 0.596705240257(8) | −0.0330(2) | 7.8 | 28.6 | 12620.2 |
| J1639−1126 | 0.698961222411(11) | −0.11834(11) | 32.7 | 9.4 | 18838.1 |
| J1647+6609 | 0.625078358705(3) | −0.306617(3) | 75.7 | 3.2 | 35854.7 |
| J1741−2152 | 0.38977960839(8) | −0.0194(8) | 3.0 | 31.9 | 18304.1 |
| J1819+0322 | 1.25130512248(2) | −0.0199(3) | 9.8 | 99.5 | 3227.2 |
| J1930+6205 | 0.686758970941(3) | −0.078006(3) | 21.1 | 13.9 | 15703.9 |
| J1948−2730 | 3.01539684110(3) | −0.026(2) | 30.7 | 185.3 | 981.2 |
| J1954+4347 | 0.720958989860(3) | −0.114606(2) | 32.6 | 10.0 | 17696.5 |
| J2018−0414 | 24.62313693500(3) | −0.00746(7) | 72.5 | 5229.2 | 22.6 |
| J2023+0937 | 0.62315481163(2) | −0.0289(2) | 7.1 | 34.2 | 11060.3 |
| J2029+5459[a] | 1.727386681(3) | −0.673(2) | 458.8 | 4.1 | 11561.0 |
| J2038+3447[a] | 6.24356709530677(95) | −0.0035100(12) | 8.7 | 2818.3 | 121.5 |
| J2040−2156 | 1.77753964687(2) | −0.0415(13) | 29.1 | 67.8 | 2751.0 |
| J2104+2830 | 2.464698618670(6) | −0.059259(5) | 57.7 | 65.9 | 2013.2 |
| J2145+2158 | 0.704725369840(4) | −0.110352(4) | 30.7 | 10.1 | 17968.4 |
| J2158−2734 | 2.09556676019757(99) | −0.03260(12) | 27.0 | 101.8 | 1904.7 |
| J2202+5040[a] | 1.34159887359(6) | −0.799249(13) | 423.3 | 2.7 | 18410.1 |
| J2210+5712 | 0.487055824584(3) | −0.044886(3) | 8.6 | 17.2 | 19945.2 |
| J2214+5357[a] | 1.331212824278(3) | −0.107256(3) | 56.4 | 19.7 | 6823.2 |
| J2242+6346[a] | 2.16939642526(3) | −2.515(2) | 2153.8 | 1.4 | 15881.8 |
| J2252+2455 | 0.5562019181950(7) | −0.034476(7) | 7.6 | 25.6 | 14323.9 |
| J2316+5619[a] | 0.941974122159(14) | −0.1119377(99) | 41.6 | 13.3 | 11710.6 |
| J2326+6243 | 3.75729134993(8) | −3.62843(2) | 5382.1 | 1.6 | 8369.5 |
| J2351+6500[a] | 0.8584551911(2) | −2.06348(3) | 699.3 | 0.7 | 57792.8 |

**Note.** Values in parentheses are the 1σ uncertainty in the last digit as reported by PINT.
[a] Pulsars discovered in Hessels et al. (2008).

### 4.2.2. J0212+5222

The timing residuals for PSR J0212+5222 developed similarly to those of PSR J1923+4243: solutions after ≃59200 required an increasing number of spin derivatives and still showed some signature in the TOAs. We replaced the derivatives (third order and above, maintaining the first and second) with a glitch model as described for J1923+4243, and the model did improve. However, the residuals did not become decisively flat as they had for the prior source. In particular, with the improved fit, another glitch-like event became evident in the archival GBT observations. We added a second glitch component and refit only archival data; we measured a second glitch with reasonable magnitude close to MJD 56551 (F-test comparison before/after this glitch in archival data returned $10^{-20}$). We did not measure any significant recovery terms for this glitch. Including the magnitude and epoch parameters in the model when we fit all TOAs flattened the residuals. The residuals before and after adding the glitches are included in Figure 3, and the parameters for the glitches are provided in Table 3.

### 4.2.3. J2029+5459

As with the other glitching sources, fits to PSR J2029+5459's timing residuals using standard timing parameters were insufficient in removing structure from the TOAs. Unlike many other sources, the phase JUMP between archival GBT data and CHIME/Pulsar observations was not enough to solve the source, and so we focused on the higher-cadence CHIME/Pulsar observations. A glitch was inserted near MJD 59536, and the CHIME/Pulsar TOAs flattened. However, connection to the archival data was still not possible, and it appeared as though there was a discontinuity in spin frequency between the two observing epochs that was too large to be due to quadratic spin-down. A second glitch was inserted in the gap (MJD 58000) and phase connection was made possible. Due to the lack of data, the epoch of this glitch is entirely covariant with the GBT–CHIME jump, and so we are unable to precisely measure it. The F-test comparing the pre- and post-glitch fits returned $\leqslant 10^{-230}$. Residuals for the data are shown in Figure 4, and the measured parameters are included in Table 3.

Table 2
Position Parameters for New GBNCC Survey Discoveries

| PSR | Longitude (deg) | Latitude (deg) | α (deg) | δ (deg) | $d_{NE}$ (kpc) | $d_{YMW}$ (kpc) |
|---|---|---|---|---|---|---|
| J0032+6946 | 50.71621375(8) | 57.28423181(5) | 8.17185916(2) | 69.77445797(7) | 2.8 | 2.3 |
| J0054+6650[a] | 49.753944(8) | 53.746402(5) | 13.7309529(2) | 66.839938(7) | 1.0 | 0.9 |
| J0110−2223 | 6.75994(12) | −27.42791(12) | 17.56944(5) | −22.3995(2) | 1.0 | 25.0 |
| J0120+1837 | 25.529881(7) | 9.43371(4) | 20.050086(9) | 18.63253(4) | 0.6 | 1.0 |
| J0141+6303 | 52.5025597(6) | 47.7773462(5) | 25.44066774(11) | 63.0637344(7) | 44.3 | 8.8 |
| J0406+3039 | 65.51429844(8) | 9.6019364(5) | 61.63549335(3) | 30.6617779(5) | 1.7 | 1.6 |
| J0415+6111 | 74.168570(14) | 39.18162(2) | 63.965032(12) | 61.19773(2) | 2.3 | 1.8 |
| J0420+4451[a] | 71.064362(4) | 23.035516(9) | 65.087665(3) | 44.850692(9) | 2.1 | 1.7 |
| J0530−3847 | 77.59187(7) | −61.90318(4) | 82.54063(3) | −38.78482(5) | 47.5 | 25.0 |
| J0749+5720 | 107.7939243(12) | 35.590651(2) | 117.425513(2) | 57.3478843(14) | 1.0 | 1.6 |
| J0758−3002 | 131.02797(4) | −49.34832(2) | 119.60286(3) | −30.039049(7) | 0.6 | 0.4 |
| J1327+3423 | 184.19721728(3) | 39.79648991(3) | 201.78145168(4) | 34.39379904(2) | 0.5 | 0.3 |
| J1354+2453 | 196.126270(12) | 33.966657(15) | 208.55576(2) | 24.896155(9) | 25.8 | 25.0 |
| J1602−1009 | 240.53824969(4) | 10.3230616(3) | 240.55533875(9) | −10.1555294(2) | 1.6 | 2.4 |
| J1604−0057 | 239.19070(3) | 19.43743(11) | 241.11460(5) | −0.96563(11) | 1.6 | 3.1 |
| J1639−1126 | 250.04038(3) | 10.6267(2) | 249.98297(6) | −11.4363(2) | 50.0 | 6.3 |
| J1647+6609 | 174.98236(6) | 82.746773(9) | 251.88529(2) | 66.139462(5) | 1.3 | 2.1 |
| J1741−2152 | 265.8115(3) | 1.50(2) | 265.4873(9) | −21.88(2) | 3.2 | 4.6 |
| J1819+0322 | 275.38135(5) | 26.72327(9) | 274.81349(4) | 3.37386(9) | 3.2 | 5.8 |
| J1930+6205 | 346.08233(5) | 79.285930(11) | 292.676907(7) | 62.092070(15) | 5.7 | 10.7 |
| J1948−2730 | 294.11015(3) | −6.3255(5) | 297.24377(6) | −27.5130(5) | 2.1 | 6.8 |
| J1954+4347 | 318.69175(2) | 62.638821(8) | 298.660222(6) | 43.960397(10) | 7.1 | 10.1 |
| J2018−0414 | 305.834950(2) | 15.008889(8) | 304.5433499(4) | −4.236871(8) | 1.5 | 1.8 |
| J2023+0937 | 310.87125(6) | 28.13128(13) | 305.82515(2) | 9.63248(14) | 3.2 | 5.2 |
| J2029+5459[a] | 345.06570(9) | 68.90190(4) | 307.320425(13) | 54.99265(5) | 3.2 | 3.5 |
| J2038+3447[a] | 325.953557(2) | 50.8339821(14) | 309.5889429(6) | 34.798367(2) | 3.4 | 3.8 |
| J2040−2156 | 306.740996(13) | −3.4638(5) | 310.06974(14) | −21.9383(5) | 1.0 | 1.7 |
| J2104+2830 | 329.759412(2) | 42.872175(2) | 316.1005626(6) | 28.515982(3) | 3.7 | 5.7 |
| J2145+2158 | 337.233375(11) | 33.23655(2) | 326.267771(2) | 21.96986(2) | 2.8 | 5.5 |
| J2158−2734 | 321.99966(2) | −14.21393(4) | 329.5035003(5) | −27.56030(5) | 0.9 | 2.7 |
| J2202+5040[a] | 2.308840(10) | 56.452520(6) | 330.6035045(7) | 50.670248(8) | 3.2 | 3.0 |
| J2210+5712 | 12.362112(15) | 60.538546(8) | 332.534489(2) | 57.216695(11) | 6.2 | 3.9 |
| J2214+5357[a] | 8.906721(4) | 57.723430(3) | 333.7247346(4) | 53.959427(3) | 5.2 | 3.7 |
| J2242+6346[a] | 28.09977(2) | 61.80138(3) | 340.65170(3) | 63.78231(2) | 3.1 | 2.6 |
| J2252+2455 | 355.093609(5) | 29.454280(9) | 343.0793086(2) | 24.931136(10) | 2.1 | 4.1 |
| J2316+5619[a] | 23.352536(8) | 53.631911(8) | 349.118791(3) | 56.333121(9) | 2.8 | 2.4 |
| J2319+6411[a] | 33.899703(8) | 58.908186(5) | 349.896653(3) | 64.190496(6) | 46.1 | 6.8 |
| J2326+6243 | 32.836873(7) | 57.337066(5) | 351.672924(3) | 62.722948(6) | 8.5 | 4.4 |
| J2351+6500[a] | 39.383882(10) | 56.896464(6) | 357.928837(2) | 65.013499(9) | 5.1 | 3.1 |

**Note.** Latitude and longitude are presented in the ecliptic frame; α, δ indicate equatorial coordinates. We include distance estimates from NE2001 (Cordes & Lazio 2002) and YMW+17 (Yao et al. 2017). Values in parentheses are the 1σ uncertainty in the last digit as reported by PINT.
[a] Pulsars discovered in Hessels et al. (2008).

#### 4.2.4. J2202+5040

Identifying the glitch progressed similarly to other sources; a model without a glitch included required several high-order polynomials to flatten TOAs, and was exchanged for a model with a glitch near MJD 59650. Similar to J2029+5459, this sufficiently flattened residuals in the CHIME data, but connection to archival GBT data required an additional glitch in the data gap. A marked difference in this source's solution is the sign of the glitches, which are both negative in Hz s$^{-1}$ (indicating a spin-down glitch). Residuals are shown in Figure 5 and glitch parameters are included in Table 3.

#### 4.2.5. J2351+6500

Timing residuals for this source hinted at the presence of a glitch shortly after MJD 59500, when the otherwise simple model no longer flattened the residuals. Fits with a glitch immediately resolved the problem. The measured glitch parameters (included in Table 3) are similar to those measured in PSRs J0212+5222 and J1923+4243, where the change in ν is positive (i.e., spin-up glitch) and the change in $\dot{\nu}$ is negative. This is apparent in the timing residuals (Figure 7), where the epoch of the glitch precedes a downward trend that recovers quadratically.

#### 4.2.6. Timing Noise and Unconfirmed Glitch Candidates

The vast majority of pulsars in this data set have been solved, meaning that their timing models correctly predict the arrival times of pulses extending into the future. There is a small subset of pulsars for which this is not true; in many of these cases, timing residuals contain additional structure that is not well fit by standard rotational/positional/binary parameters. In fact, data for PSR J1954+3852 can be fit with as many as four significant spin derivatives. Binary motion can manifest as apparent spin derivatives in timing data when the orbital period is much larger

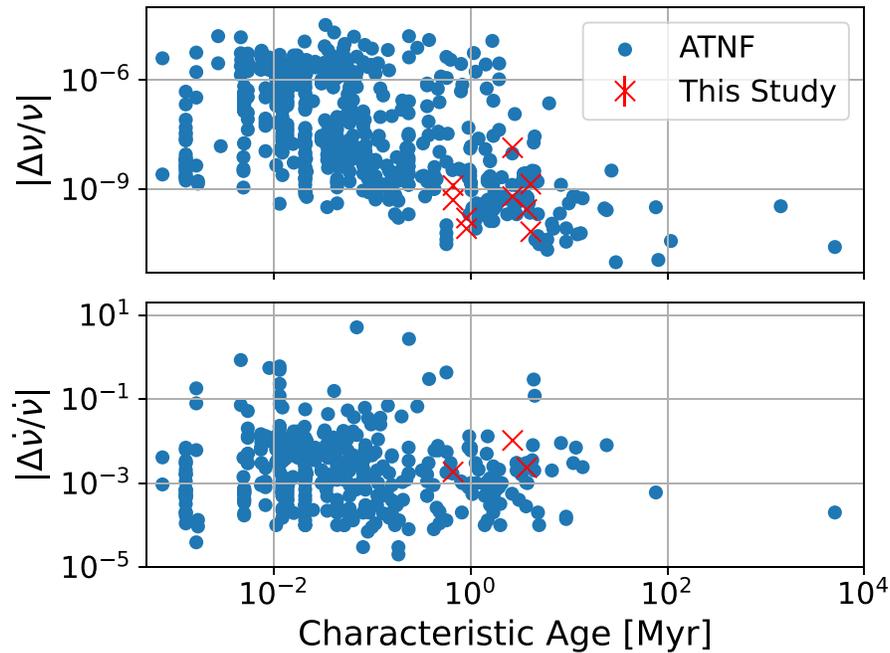

**Figure 2.** Glitch magnitude vs. characteristic age. Fractional changes in both spin frequency (top panel) and frequency derivative (bottom panel) are shown in red, while published glitches from the ATNF glitch catalog are shown in blue.

**Table 3**
Glitches in GBNCC Pulsars

| PSR | Epoch (MJD) | $\Delta f$ ($10^{-10}$ Hz) | $\Delta \dot{f}$ ($10^{-17}$ Hz s$^{-1}$) | $\Delta \ddot{f}$ ($10^{-25}$ Hz s$^{-2}$) |
|---|---|---|---|---|
| J0212+5222 | 56571(5) | 4.2(4) | … | … |
|  | 59203(4) | 2.13(7) | −0.01(5) | −2.15(11) |
| J1923+4243 | 59528(2) | 4.6(2) | −1.68(12) | 5.4(6) |
| J2029+5459 | 59531(8) | 1.06(3) | … | … |
|  | 58000[a] | 23.4(15) | … | … |
| J2202+5040 | 59657(4) | −8.4(4) | −8.3(3) | … |
|  | 58000[a] | −182.9(6) | … | … |
| J2351+6500 | 59332(4) | 4.2(2) | −3.82(4) | … |

**Notes.** Values in parentheses are the $1\sigma$ uncertainty in the last digit as reported by PINT.
[a] The second listed glitch for PSRs J2029+5459 and J2202+5040 occurred in a large data gap between GBT and CHIME/Pulsar observations; because of this, the glitch epochs are not uniquely determined. Instead, we include the epoch used in the models, which were arbitrarily chosen.

than the data span (for one such example, see Bassa et al. 2016), but no such model improves the fit. We assume that the residual structure is due to timing noise; the daily cadence at CHIME gives us unprecedented sensitivity to variations in period on short timescales. This can be quantified in a simple case where we assume the uncertainty in polynomial fits scales as $\sigma_{TOA}/\sqrt{N}$. The TOA uncertainty will depend on the telescope used and the pulsar's intrinsic brightness; given that the CHIME system's sensitivity is approximately 70% that of the 820 MHz system used for the GBT and that pulsar flux density scales with observing frequency (average spectral index of −1.4; Bates et al. 2014), we can directly compare the (analytic) uncertainties on timing noise measurement. We have assumed the pulse width does not change significantly between the observing frequencies, which is not always true. However, between 600 and 820 MHz, this change is likely negligible. GBT timing campaigns for canonical pulsars typically include roughly one observation per month with duration ⩾30 minutes. Comparing this to weekly 10 minutes CHIME scans, the GBT measurement uncertainty is approximately twice that of CHIME; to match CHIME uncertainty, ⩾100 minutes scans would be needed at GBT. Usage of a lower frequency at GBT can help this, but only for sources with steep spectral indices—and CHIME cadences/durations are generally greater than those used above.

One metric used to quantify timing noise is the stability parameter, defined as (Arzoumanian et al. 1994)

$$\Delta_8 = \log\left(\frac{1}{6\nu}|\ddot{\nu}|T_8^3\right),\quad(1)$$

where $\nu$ is the pulsar spin frequency, $\ddot{\nu}$ is its second derivative, and $T_8 = 10^8$ s is the baseline over which the former parameters are measured. This study has produced measurements of $\ddot{\nu}$ for 15 sources; for these and all other sources, we check for baselines in CHIME-only data that span at least $0.8\,T_8$ and fit for $\ddot{\nu}$ over this period. These measurements are shown in Figure 8 with our best-fit line (with $3\sigma$ uncertainty) and the line published in Arzoumanian et al. (1994) for comparison. We note that our measurements deviate from those of Arzoumanian et al. (1994) slightly, particularly for sources with larger spin-down (i.e., younger pulsars). Models that did not previously include $\ddot{\nu}$ (plotted with green triangles in Figure 8) agree more closely with those results. In general, though, the spread is fairly large.

### 4.3. Proper Motions

Measurements of pulsar proper motions over short timescales can be difficult to reliably disentangle from position, as errors in the two components will manifest as annual sinusoids. Breaking this degeneracy is possible with high cadence and long observation baselines, which we have achieved for many sources included in this study. Table 4 provides all of the sources for which we have measured proper motion. Many of these sources have been measured in previous studies; we utilize this archival data to greatly supplement our measurements. Sources for which no previous measurements exist are marked and explained in the table footnotes.

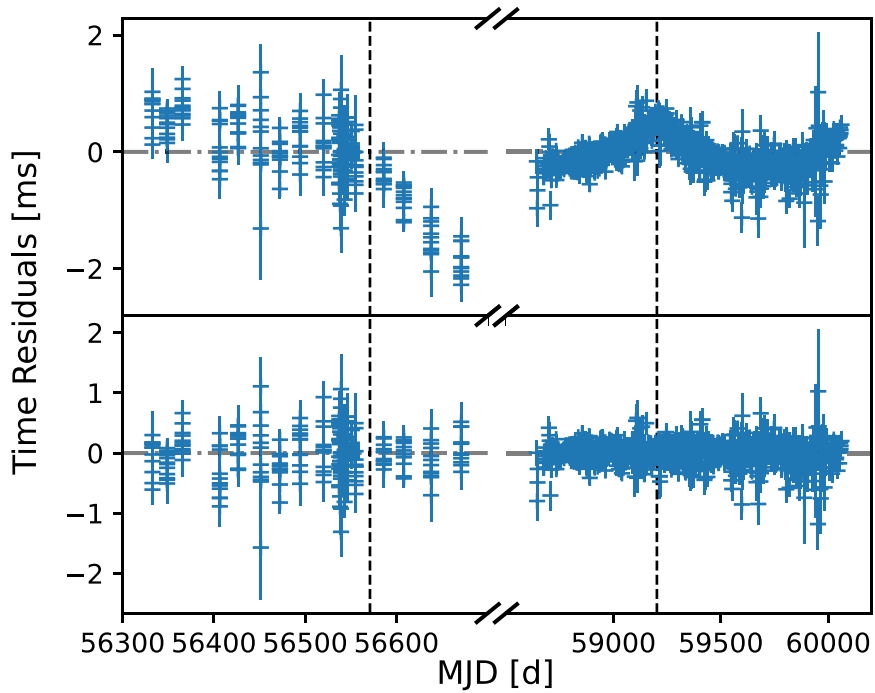

**Figure 3.** Glitches in PSR J0212+5222 timing residuals. Dashed vertical lines highlight the epochs of the two measured glitches; we use a broken horizontal axis for clarity.

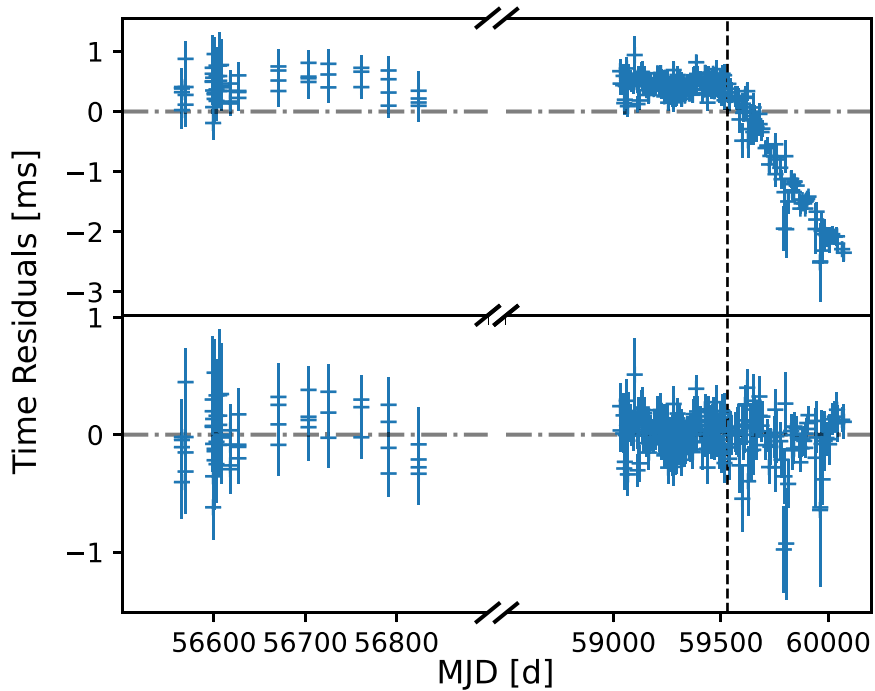

**Figure 4.** Glitch in PSR J2029+5459 timing residuals. The epoch of the glitch is marked with a vertical dashed line, and the horizontal axis is broken for clarity. For this source, we also fit a glitch in the gap between the GBT data and CHIME data. We are not able to constrain the epoch of this glitch, so it is omitted from the plot.

For each measurement of proper motion, we also estimate the pulsar's transverse velocity using the DM distance of the pulsar, which we calculate using both NE2001 (Cordes & Lazio 2002) and YMW+16 (Yao et al. 2017). With a distance and a velocity measurement, we can further refine the pulsar's rate of spin-down by accounting for two effects: Galactic acceleration and transverse motion (Shklovskii 1970). These terms can be calculated and removed from the measured period derivative of a pulsar to calculate a more accurate magnetic field and spin-down luminosity. For a full description of the procedure used to estimate these terms, see Lynch et al. (2018b). Swiggum et al. (2023) also incorporates some new measurements and tools to help with such calculations.

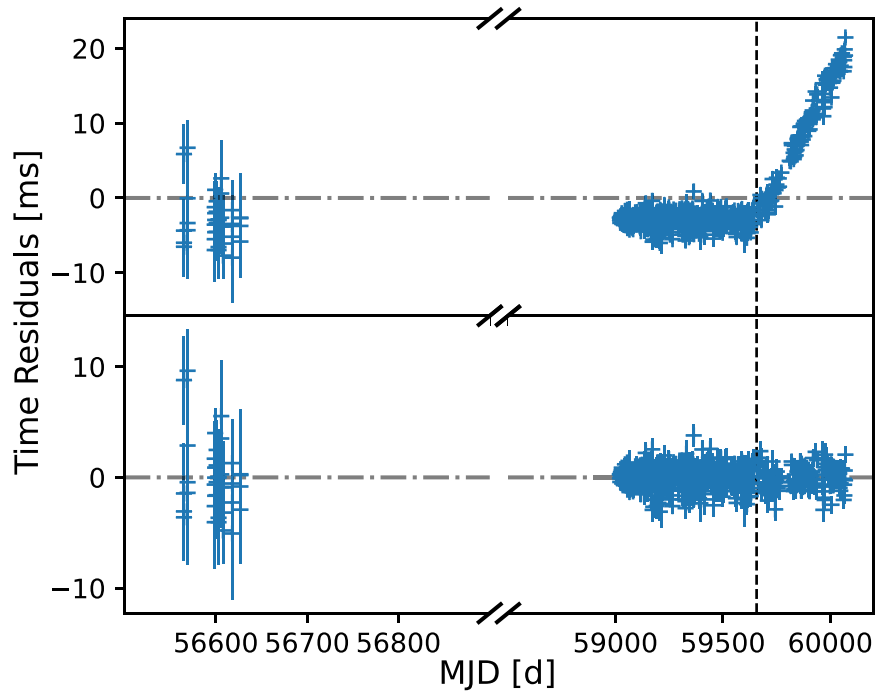

**Figure 5.** Glitch in PSR J2202+5040 timing residuals. The epoch of the glitch is marked with a vertical dashed line, and the horizontal axis is broken for clarity. For this source, we also fit a glitch in the gap between the GBT data and CHIME data. We are not able to constrain the epoch of this glitch, so it is omitted from the plot.

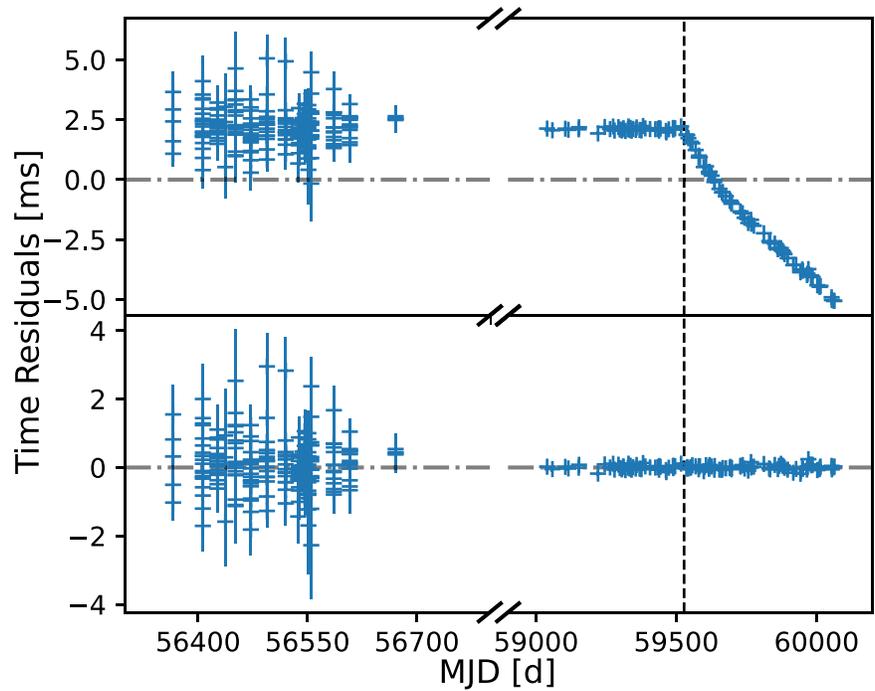

**Figure 6.** Glitch in PSR J1923+4243 timing residuals. The dashed vertical line indicates the epoch of the glitch. For clarity, the horizontal axis is broken to omit the data gap between MJD 56700 and 59000.

Of the 19 sources for which we measure proper motion, 13 are new measurements. The remaining six sources (J0645+5158, J1125+7819, J1641+8049, J1710+4923, J1816+4510, and J1955+6708) have published proper motions in the ATNF catalog (v1.70; Manchester et al. 2005). Some of these measurements differ from published values by as much as $79\sigma$, or nearly 40 mas yr$^{-1}$. We include in Figure 9 a comparison between these previous measurements and those coming from this study. PSRs J0645+5158, J1125+7819, and J1816+4510 are fairly consistent, and the other three are very disparate. PSR J1641+8049 is a spider pulsar (for an overview, see Fruchter et al. 1988), which is actively ablating its companion. These systems often exhibit variable orbital parameters due to their dynamic environment (see, e.g., Polzin et al. 2019). As noted in Mata Sánchez et al. (2023), proper-motion measurements for this source from Lynch et al. (2018b) are likely overestimated, and are at odds with measurements from optical data. This is likely what is reflected in our new measurement, as the timing baseline in this work is approximately twice that used in Lynch et al. (2018b). The timing baselines for PSRs J1955+6708 and J1710+4923 have doubled, as well, which may have resulted in the significant change.

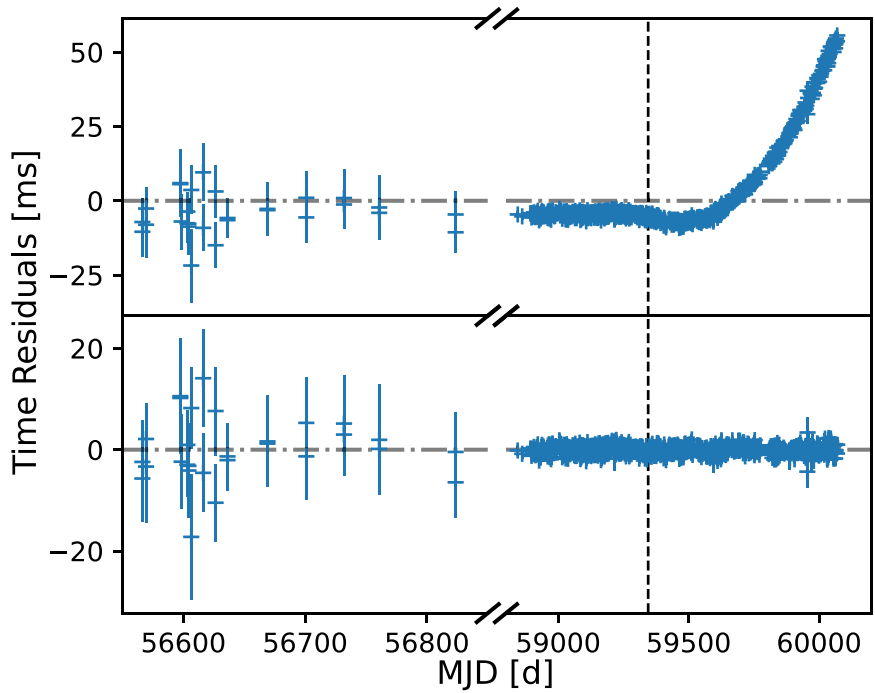

**Figure 7.** Glitch in PSR J2351+6500 timing residuals. The epoch of the glitch is marked with a vertical dashed line, and the horizontal axis is broken for clarity.

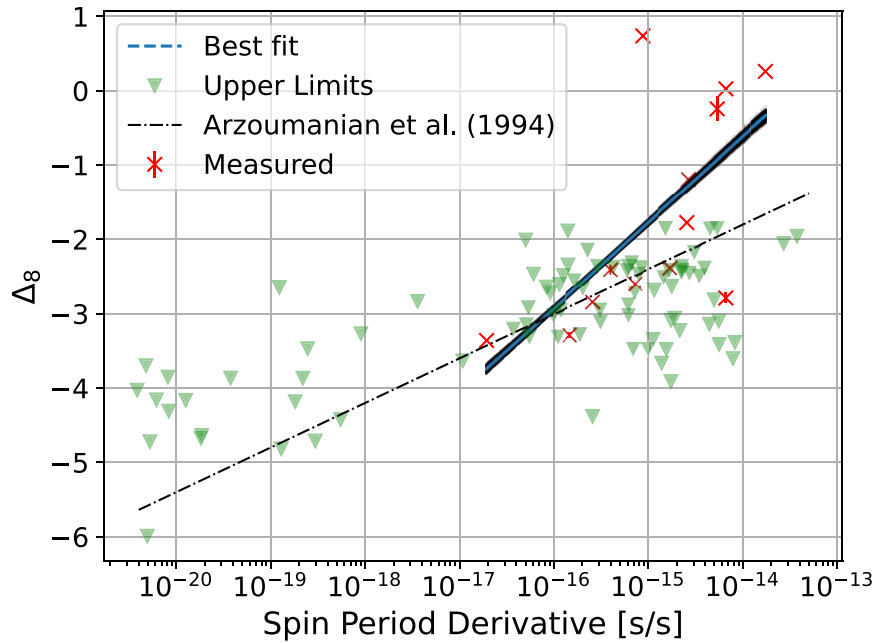

**Figure 8.** Timing stability parameter $\Delta_8$ for GBNCC sources. For each source, we limited the baseline to $\approx 10^8$ s in CHIME-only data. Red X markers indicate sources where $\ddot{\nu}$ had been incorporated in the timing model previously; green triangles indicate sources where we added $\ddot{\nu}$ to the model to determine a limit. The black dashed–dotted line is the fit line from Arzoumanian et al. (1994), and the blue dashed line is a best-fit to the red points. We also include the $3\sigma$ region for the best-fit line, which shows apparent discrepancy with the prior fit; we attribute this to the number of points observed and implicit scatter in the $\Delta_8$ parameter.

### 4.4. Binary Pulsars

The GBNCC survey has been an excellent tool for finding binary pulsars, with a total of 24 such discoveries. Here, we have included long-term timing solutions for 18 of its binaries, plus an additional binary pulsar from the GBT350 survey (Hessels et al. 2008). These sources span a wide range of orbital periods ($0.1 \leqslant P_B \leqslant 528$ days) and eccentricities ($10^{-6} \leqslant e \leqslant 0.6$), encompassing many evolutionary pathways and timing phenomena. We plot each binary system's orbital period and minimum companion mass in Figure 10. Tables 5 and 6 include the measured/derived binary parameters for all sources included in this data set. The first includes sources which use the ELL1 binary model (Lange et al. 2001), a parameterization that is appropriate for sources in nearly

Table 4
Proper Motion Measurements for GBNCC Sources

| PSR | $\mu_\lambda$ (mas yr$^{-1}$) | $\mu_\beta$ (mas yr$^{-1}$) | $d$ (kpc) | $v_{\rm tot}$ (km s$^{-1}$) | $\dot{P}_{\rm Gal}$ ($10^{-21}$ s s$^{-1}$) | $\dot{P}_{\rm S}$ ($10^{-21}$ s s$^{-1}$) | $\dot{P}_{\rm int}$ ($10^{-21}$ s s$^{-1}$) | B ($10^8$ G) | $\tau_{\rm c}$ (Gyr) | $\dot{E}$ ($10^{33}$ erg s$^{-1}$) |
|---|---|---|---|---|---|---|---|---|---|---|
| J0214+5222[a] | 11.6(3) | 4.5(5) | 1.0 | 60.8 | −0.04 | 9.52 | 286.92 | 26.9 | 1.4 | 0.8 |
|  |  |  | 1.2 | 68.4 | −0.02 | 10.71 | 285.71 | 26.8 | 1.4 | 0.8 |
| J0406+3039[a] | 2.37(8) | −4.6(5) | 1.7 | 42.9 | 0.18 | 0.30 | 7.82 | 1.4 | 5.3 | 17.4 |
|  |  |  | 1.6 | 40.1 | 0.17 | 0.28 | 7.86 | 1.4 | 5.3 | 17.5 |
| J0509+3801[a] | 3.1(3) | −7(2) | 1.9 | 66.6 | 10.17 | 19.07 | 7905.96 | 248.9 | 0.2 | 0.7 |
|  |  |  | 1.6 | 54.0 | 8.53 | 15.47 | 7911.19 | 249.0 | 0.2 | 0.7 |
| J0645+5158 | 0.80(2) | −7.54(4) | 0.7 | 25.2 | 0.10 | 0.87 | 3.96 | 1.9 | 35.4 | 0.2 |
|  |  |  | 0.7 | 24.1 | 0.09 | 0.83 | 4.01 | 1.9 | 35.0 | 0.2 |
| J0742+4110[a] | −8.47(12) | −12.7(6) | 0.7 | 51.8 | −0.01 | 1.27 | 5.43 | 1.3 | 9.2 | 6.9 |
|  |  |  | 0.5 | 37.8 | −0.02 | 0.93 | 5.78 | 1.4 | 8.6 | 7.4 |
| J1125+7819 | 16.151(13) | 22.22(2) | 0.6 | 82.4 | −0.36 | 4.87 | 2.43 | 1.0 | 27.4 | 1.3 |
|  |  |  | 0.8 | 108.6 | −0.41 | 6.42 | 0.94 | 0.6 | 70.8 | 0.5 |
| J1221−0633[a] | −0.56(12) | −6(2) | 0.7 | 21.0 | −0.29 | 0.12 | 5.42 | 1.0 | 5.7 | 29.6 |
|  |  |  | 1.3 | 35.2 | −0.38 | 0.21 | 5.42 | 1.0 | 5.7 | 29.6 |
| J1327+3423[a] | −8.49(3) | 1.44(4) | 0.5 | 19.4 | −5.84 | 3.56 | 131.58 | 23.7 | 5.0 | 0.1 |
|  |  |  | 0.3 | 14.0 | −4.92 | 2.57 | 131.65 | 23.7 | 5.0 | 0.1 |
| J1434+7257[a] | 7.31(14) | −7.1(2) | 0.7 | 34.4 | −4.96 | 7.49 | 545.83 | 48.3 | 1.2 | 0.3 |
|  |  |  | 1.0 | 47.0 | −5.87 | 10.22 | 544.00 | 48.2 | 1.2 | 0.3 |
| J1628+4406[a] | −0.5(2) | −20.3(2) | 0.6 | 58.8 | −21.00 | 111.00 | 19238.61 | 597.4 | 0.1 | 0.1 |
|  |  |  | 0.5 | 52.0 | −19.46 | 98.30 | 19249.77 | 597.6 | 0.1 | 0.1 |
| J1641+8049 | 1.50(2) | −0.14(2) | 1.7 | 11.8 | −0.31 | 0.02 | 10.07 | 1.4 | 3.2 | 48.2 |
|  |  |  | 3.0 | 21.6 | −0.42 | 0.03 | 10.17 | 1.5 | 3.1 | 48.6 |
| J1710+4923[b] | −49.65(4) | −44.25(4) | 0.7 | 207.8 | −0.36 | 22.81 | … | … | … | … |
|  |  |  | 0.5 | 159.6 | −0.30 | 17.51 | 0.99 | 0.6 | 51.7 | 1.2 |
| J1806+2819[a] | −1.7(6) | −11.6(9) | 1.3 | 74.1 | −1.59 | 6.69 | 32.45 | 7.1 | 7.4 | 0.4 |
|  |  |  | 1.3 | 73.6 | −1.58 | 6.64 | 32.49 | 7.1 | 7.4 | 0.4 |
| J1816+4510 | 0.07(2) | −4.60(2) | 2.4 | 52.8 | −0.84 | 0.40 | 43.52 | 3.8 | 1.2 | 52.8 |
|  |  |  | 4.4 | 95.1 | −1.42 | 0.72 | 43.78 | 3.8 | 1.2 | 53.1 |
| J1938+6604[a] | 2.95(10) | 0.67(8) | 2.3 | 33.5 | −4.76 | 1.16 | 22.14 | 7.1 | 15.9 | 0.1 |
|  |  |  | 3.4 | 48.5 | −5.99 | 1.68 | 22.86 | 7.2 | 15.4 | 0.1 |
| J1955+6708 | −4.46(14) | 3.64(13) | 3.4 | 92.3 | −2.19 | 2.33 | 12.44 | 3.3 | 10.9 | 0.8 |
|  |  |  | 5.3 | 144.8 | −2.69 | 3.66 | 11.61 | 3.2 | 11.7 | 0.7 |
| J2022+2534[a] | −5.42(7) | −4.49(10) | 3.3 | 110.6 | −0.83 | 1.06 | 5.93 | 1.3 | 7.1 | 12.6 |
|  |  |  | 4.0 | 134.0 | −1.03 | 1.28 | 5.91 | 1.3 | 7.1 | 12.6 |
| J2123+5434[a] | −13.7(7) | 4.9(7) | 2.1 | 144.3 | −23.05 | 149.80 | 91.90 | 36.2 | 23.9 | $1.6 \times 10^{-3}$ |
|  |  |  | 1.8 | 125.2 | −20.49 | 130.01 | 109.12 | 39.4 | 20.2 | $1.4 \times 10^{-3}$ |
| J2150−0326[a] | 1.60(13) | −11.12(96) | 1.1 | 56.2 | −0.46 | 1.14 | 7.48 | 1.6 | 7.4 | 6.8 |
|  |  |  | 2.0 | 105.2 | −0.68 | 2.13 | 6.71 | 1.6 | 8.3 | 6.1 |

**Notes.** Pulsar proper motions are given in the ecliptic frame. For each source, we calculate quoted parameters using distance predictions from both NE2001 (Cordes & Lazio 2002) and YMW+16 (Yao et al. 2017) and estimate the Galactic potential using results from Guo et al. (2021). Additional parameters include the total velocity $v_{\rm tot}$, apparent spin derivatives from Galactic motion $\dot{P}_{\rm Gal}$ and Shklovskii $\dot{P}_{\rm S}$ corrections, the remaining intrinsic spin-down $\dot{P}_{\rm int}$, magnetic field $B$, characteristic age $\tau_{\rm c}$, and spin-down luminosity $\dot{E}$. Values in parentheses are the $1\sigma$ uncertainty in the last digit as reported by PINT. Uncertainties on distances are generally $\simeq 30\%$.
[a] New measurements of proper motion.
[b] Proper-motion measurements for PSR J1710+4923 imply a negative intrinsic spin-down. This is most likely due to an error in the distance estimate and leads to an erroneous magnetic field and characteristic age. For this reason, we omit terms for this source from the final three columns. Note that this was also mentioned in Lynch et al. (2018b).

circular orbits. The second includes sources that use either the more generalized binary model (BT; Blandford & Teukolsky 1976) or the relativistic parameterization (DD; Damour & Deruelle 1985, 1986) for sources with significant pulsar acceleration. Many of these binary systems have been published elsewhere, though few data sets have the long baselines/high cadences reported here. We include parameters for nine new binary solutions, and focus on several sources of particular interest below.

### 4.4.1. Post-Keplerian Parameter Measurements

Interestingly, despite the dramatic increase in timing baselines for the pulsars in this study, only one post-Keplerian term has been added to our binary models: the orbital decay term $\dot{P}_{\rm B}$ for PSR J0509+3801, which we measure to be $-1.37(7) \times 10^{-12}$ s s$^{-1}$. For all binaries, we calculated estimates of these parameters using the equations 8.48–8.52 in Lorimer & Kramer (2012) and determined how these the inclusion of these estimates would impact our timing solutions. Aside from the few sources with post-Keplerian measurements from previous works (see below), none of these parameters are predicted to be measured significantly. We also added them to the fit parameters as a test, and no measurements were significant above $2\sigma$.

PSRs J0509+3801 (Lynch et al. 2018b) and J1759+5036 (Agazie et al. 2021) are both in relativistic orbits with prior measurements of the advance of periastron $\dot{\omega}_{\rm orbit}$—the former also has a previous measurement of the Einstein delay parameter $\gamma$. We improve previous constraints on these parameters, and these improvements are included in Table 7. The improvement in precision for PSR J0509+3801's binary parameters also further constrain the masses in the orbit (previously published in Lynch et al. 2018b), measured to be 1.399(6) $M_\odot$ and 1.412(6) $M_\odot$, respectively, for the pulsar and its companion (shown in Figure 11). All mass measurements are consistent with those made in previous works, including the new $\dot{P}_B$ measurement (though it is not yet a strong constraint compared to those made from $\dot{\omega}$ and $\gamma$). For PSR J1759+5036, our $\dot{\omega}_{\rm orbit}$ measurement of 0.1289(4) deg yr$^{-1}$ is more constraining than the previously published value (0.127(10) deg yr$^{-1}$; Agazie et al. 2021). This further constrains the total system mass to 2.679(12) $M_\odot$, which agrees within $2\sigma$ of the mass published in Agazie et al. (2021). Considering that the timing baseline for this source has approximately doubled with the inclusion of CHIME observations, such a change is not unexpected.

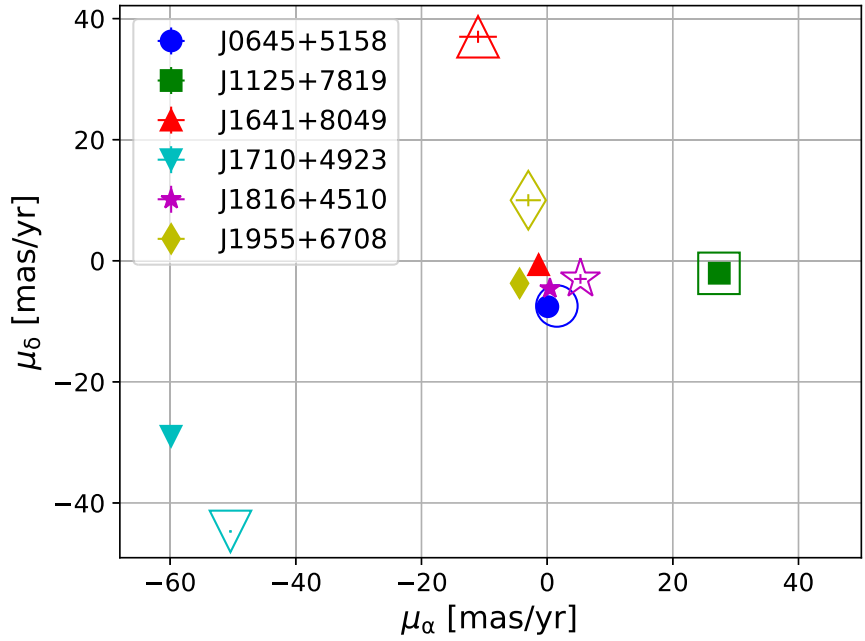

**Figure 9.** Comparison of proper-motion measurements to previously published values. For each of the six sources with measurements of proper motion, we plot a hollow marker for the previous value and a filled marker for our measurement. All points have error bars, but they are small in most cases. The dramatic change in proper motion for PSR J1641+8049 is supported by optical results published by Mata Sánchez et al. (2023).

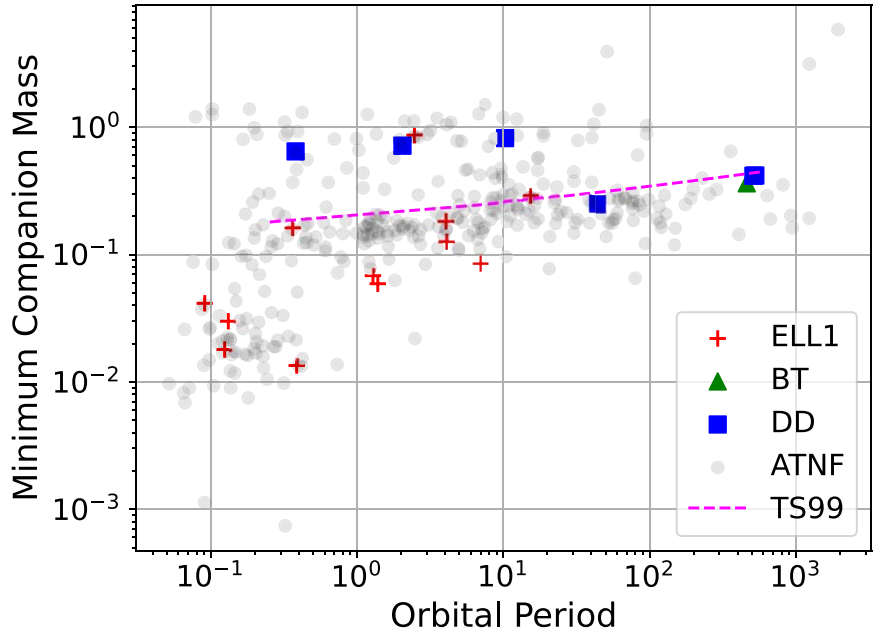

**Figure 10.** Orbital period vs. companion mass. Bold points indicate pulsars in our sample, and fainter points come from the ATNF catalog. Marker type indicates what binary model is used. The dashed magenta line shows the relationship between white dwarf mass and orbital period given in Tauris & Savonije (1999).

### 4.4.2. J2038+3447

PSR J2038+3447 (J2038) was originally discovered in the NGP survey (Hessels et al. 2008) and subsequently observed in 820 MHz observations that spanned $\simeq$226 days. The first of these observations included gridding to localize the pulsar to within an

Table 5
Binary Pulsars that Utilize ELL1 Models

| PSR | Measured | | | | | Derived | |
|---|---|---|---|---|---|---|---|
| | $P_B$ (days) | $a \sin i/c$ (s) | $T_{asc}$ (MJD) | $\epsilon_1$ ($10^{-5}$) | $\epsilon_2$ ($10^{-5}$) | $f_M$ ($10^{-3} M_\odot$) | $M_{c,min}$ ($M_\odot$) |
| J0406+3039[b] | 6.955717217(2) | 2.3192820(2) | 57451.8533146(4) | 0.37(2) | −1.17(2) | 0.27686 | 0.08 |
| J0742+4110[a] | 1.3853611813(6) | 0.5564571(11) | 56045.1468648(13) | 0.1(4) | −0.0(4) | 0.096394 | 0.06 |
| J1125+7819 | 15.3554459628(3) | 12.19242105(9) | 56157.47634435(5) | −1.2856(15) | 0.083(2) | 8.2533 | 0.29 |
| J1221−0633[a] | 0.3863495385(4) | 0.0552868(5) | 57906.1230476(11) | 23(2) | 4(2) | 0.0012156 | 0.01 |
| J1239+3239[a] | 4.0854016578(12) | 2.3711240(10) | 59282.5326304(3) | 0.70(9) | 0.20(9) | 0.85758 | 0.13 |
| J1602−1009[b] | 0.12445961397(13) | 0.0346079(3) | 58988.1081132(2) | 20(2) | −4(2) | 0.0028731 | 0.02 |
| J1720−0534 | 0.1316985759(2) | 0.0596156(9) | 59144.1417331(3) | 2(3) | −2(2) | 0.013116 | 0.03 |
| J1816+4510 | 0.36089348174(2) | 0.5954001(2) | 55771.10465888(2) | 0.44(6) | −0.16(6) | 1.7400 | 0.16 |
| J1816+4510 | 0.360893481832(4) | 0.5953998(2) | 55771.10465783(3) | 0.41(5) | −0.07(5) | 1.7400 | 0.16 |
| J1938+6604 | 2.46716272965(6) | 8.9507426(6) | 56366.09670039(6) | 0.585(13) | −2.735(13) | 126.49 | 0.87 |
| J2022+2534[a] | 1.2837028285(2) | 0.6092408(3) | 57535.5638692(3) | 0.105(98) | 0.14(9) | 0.14734 | 0.07 |
| J2150−0326[a] | 4.0445506010(5) | 3.3207135(6) | 57510.6590599(2) | 0.37(3) | 0.95(3) | 2.4035 | 0.18 |

**Notes.** All timing models presented here use the ELL1 binary model, which is appropriate for low-eccentricity orbits. Values in parentheses are the 1$\sigma$ uncertainty in the last digit as reported by PINT.
[a] Published rotational models, but no previously measured binary parameters.
[b] Not published prior to this work.

Table 6
Binary Pulsars that Utilize DD/BT Models

| PSR | Measured | | | | | Derived | |
|---|---|---|---|---|---|---|---|
| | $P_B$ (days) | $a \sin i/c$ (s) | $T_0$ (MJD) | $\omega$ (°) | $e$ | $f_M$ ($10^{-2} M_\odot$) | $M_{c,min}$ ($M_\odot$) |
| J0032+6946 | 527.62131333(13) | 178.6747603(4) | 57142.9728(6) | 147.5262(4) | 0.000531747(4) | 2.2000 | 0.42 |
| J0214+5222 | 512.0397765(5) | 174.565753(2) | 56126.6029(4) | 210.5889(3) | 0.00532803(3) | 2.1785 | 0.42 |
| J0509+3801 | 0.37958378732(9) | 2.05046(3) | 58054.1829999(4) | 144.2024(7) | 0.586409(3) | 6.4242 | 0.65 |
| J1045−0436[a] | 10.27364598(2) | 22.252634(7) | 57481.373(13) | 321.9(4) | 7.46(8)×$10^{-5}$ | 11.209 | 0.82 |
| J1759+5036 | 2.042983903(14) | 6.82456(2) | 57604.492222(11) | 92.132(2) | 0.308269(4) | 8.1767 | 0.72 |
| J1806+2819 | 43.86695922(5) | 21.608786(3) | 57028.28(2) | 257.5(2) | 8.68(2)×$10^{-5}$ | 0.56299 | 0.25 |
| J2038+3447[a] | 461.362059(3) | 144.727612(9) | 58991.3045(99) | 70.212(8) | 0.00081673(10) | 1.5292 | 0.36 |

**Note.** All timing models presented here use DD/BT binary models. Values in parentheses are the 1$\sigma$ uncertainty in the last digit as reported by PINT.
[a] Published rotational models, but no previously measured binary parameters.

Table 7
Updates to Post-Keplerian Parameters

| PSR | Parameter | Measurement |
|---|---|---|
| J0509+3801 | Advance of periastron $\dot{\omega}_{orbit}$ (deg yr$^{-1}$) | 3.03476(11) |
| | Einstein delay $\gamma$ (s) | 0.00444(3) |
| | Orbital decay $\dot{P}_B$ ($10^{-12}$) | −1.37(7) |
| | Pulsar mass, derived ($M_\odot$) | 1.399(6) |
| | Companion mass, derived ($M_\odot$) | 1.412(6) |
| | Total mass, derived ($M_\odot$) | 2.81071(14) |
| J1759+5036 | Advance of periastron $\dot{\omega}_{orbit}$ (mas yr$^{-1}$) | 0.1289(4) |
| | Total mass, derived ($M_\odot$) | 2.679(12) |

**Note.** Values in parentheses are the 1$\sigma$ uncertainty in the last digit as reported by PINT.

820 MHz beam. Subsequent observations were centered on the improved position at a varying cadence (generally in groups with a few weeks between groups). Timing analysis of these data uncovered a significant period drift between observations much larger than expected from spin-down, suggesting binary motion. However, the data span was insufficient to fully disentangle its rotational, positional, and binary parameters. Further 820 and 350 MHz observations with the GBT were conducted in late 2019/early 2020, but phase connection was still not reached.

J2038 was among the first sources to be added to the CHIME/Pulsar regular observations following the CHIME/GBNCC data sharing agreement, and has since been observed with near-daily cadence. Given the many observations and lack of binary constraint, we assumed the binary period was ≫1 yr. To handle this, we opted to include measurements of frequency derivatives in our model to capture the orbital motion (for a similar case, see Kaplan et al. 2016 and Bassa et al. 2016). These apparent period changes are actually due to binary-induced Doppler shifts in the pulsar signal, and can therefore be mapped back onto the orbital phase/orbital period plane. After ≃100 days of CHIME/Pulsar observations, this technique allowed us to predict a binary model that was phase-connected with the archival GBT data. Further observations broke covariance between position and spin-down and fully solved the system.

Determination of these parameters showed that J2038 is a relatively slow rotator ($P \simeq 160$ ms) in a large, nearly circular orbit ($P_B \simeq 462$ days, $A_1 \simeq 142$ light seconds, $e < 10^{-3}$). These measurements place it in a somewhat sparse region of the orbital period/spin period plane, as we show in Figure 12. Nearby binaries are either in highly eccentric orbits with main-sequence companions or circular orbits with degenerate (mostly carbon-oxygen or helium white dwarf) companions. Given J2038's low eccentricity and its minimum/median mass of 0.36/0.43 $M_\odot$ (calculated from the binary mass function), the companion is most likely to be a He-core white dwarf.

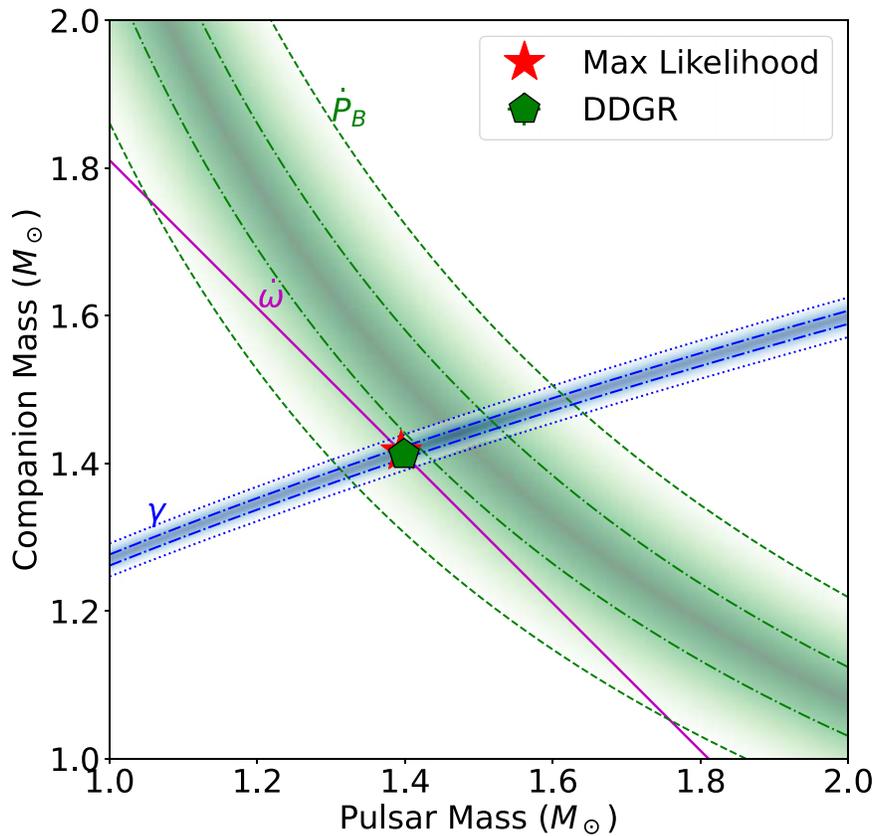

**Figure 11.** Mass–mass diagram for PSR J0509+3801. With three measurements of post-Keplerian parameters (advance of periastron, Einstein delay, and orbital decay), the masses of the pulsar and its companion are measured to be 1.399(6) $M_\odot$ and 1.412(6) $M_\odot$. We include the $3\sigma$ error regions ($3\sigma$ and $1\sigma$ lines are also shown) for each parameter and highlight the maximum likelihood with a red star. The green pentagon shows the result when fitting for the masses using a model described in Taylor & Weisberg (1989), which assumes that general relativity (GR) is the correct theory of gravity (DDGR). The new measurement of $\dot{P}_B = -1.37(7) \times 10^{-12}$ does not strongly constrain the masses, but is consistent with GR predictions for orbital decay due to the emission of gravitational waves. The uncertainty on $\dot{P}_B$ is larger than predicted contributions from Galactic/Shklovskii terms; regardless, we include these corrections in our estimates using the DM distance of 1.6 kpc. All curves are plotted under the assumption that GR is correct.

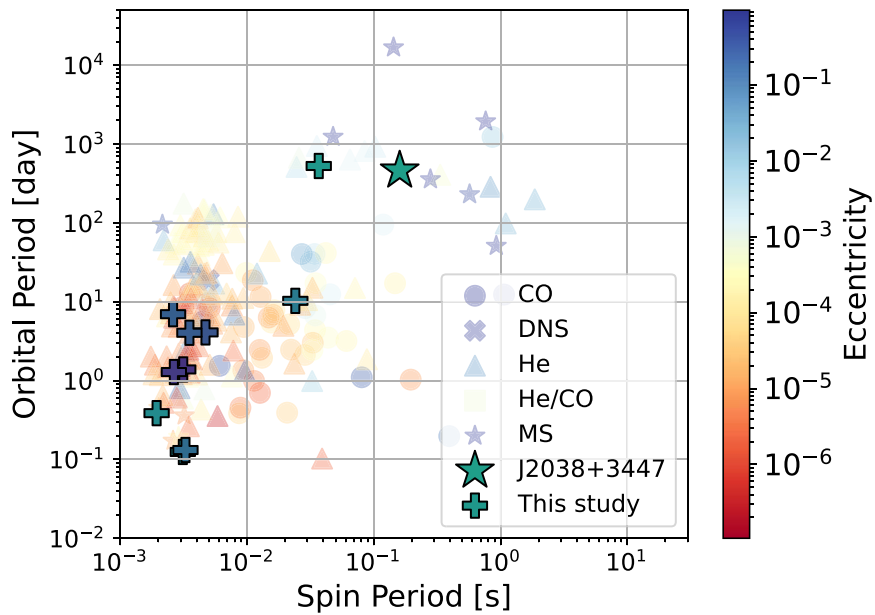

**Figure 12.** Comparison of PSR J2038+3447's spin and orbital periods to other known binaries. We include points for carbon-oxygen companions (CO, circles), double neutron star companions (DNS, filled X's), helium/helium-CO companions (He and He/CO, triangles and squares), and main-sequence companions (MS, stars). The filled and outlined star indicates J2038. Filled plus sign markers highlight the other binaries published in this paper. For all points, colors indicate the orbital eccentricity.

### 5. Conclusion

In this work, we have analyzed timing data for 128 sources discovered in GBNCC and the GBT NGP surveys. When available, we incorporate both archival data from GBT, Arecibo, and LOFAR with recent high-cadence CHIME/Pulsar data. We have connected TOA-generation and pulsar-timing tools to produce a user-friendly, self-auditing pipeline for managing large and disjoint baselines. Using these data, we fit for spin and position parameters for all sources, including 42 newly published timing solutions. Timing residuals for all sources are provided in Appendix 5. For 19 sources (13 of which are new), we measure proper motion and infer kinematic corrections to the observed spin-down. The data set has also provided measurements of binary parameters for 19 sources and has uncovered seven glitches in four pulsars. We briefly analyzed timing noise in the population, though we leave a more involved analysis for a future work. We provide timing models, TOAs, and configuration files for reproduction of results presented here on the GBNCC GitHub page.[32]

The use of CHIME follow-up data for the collection of GBNCC-discovered pulsars has provided a new probe into the short-timescale variability of their timing models. Preceding timing campaigns were generally disparate in time, often with weeks to years between subsequent observations (if timing follow-up had been conducted at all). This limits the ability to measure even the "primary" model parameters (i.e., spin, spin-down, and position), as phase connection can easily be lost between subsequent observations when the gap is large enough. In this way, high-cadence CHIME observations help to reach a stable solution more quickly than episodic GBT scans. For young pulsars, measurements of these parameters can be contaminated with glitches and timing noise; again, this is mitigated with high-cadence observations where glitches are visible in the timing residuals.

For those sources with long baselines, connection to CHIME data across large gaps (and between different observatories) signals that the timing model in use is robust. This requires very accurate timing positions and proper motions (which we see in Section 4.3) and a complete binary solution where applicable (Section 4.4). Hence, CHIME's timing capabilities improve upon previous timing campaigns by facilitating this phase connection.

### Acknowledgments

The authors thank the anonymous referee for their helpful guidance. The Green Bank Observatory is a facility of the National Science Foundation (NSF) operated under cooperative agreement by Associated Universities, Inc. The National Radio Astronomy Observatory is a facility of the NSF operated under cooperative agreement by Associated Universities, Inc. The Arecibo Observatory is a facility of the NSF operated under cooperative agreement (AST-1744119) by the University of Central Florida (UCF) in alliance with Universidad Ana G. Méndez (UAGM) and Yang Enterprises (YEI), Inc.

We acknowledge that CHIME is located on the traditional, ancestral, and unceded territory of the Syilx/Okanagan people. We are grateful to the staff of the Dominion Radio Astrophysical Observatory, which is operated by the National Research Council of Canada. CHIME is funded by a grant from the Canada Foundation for Innovation (CFI) 2012 Leading Edge Fund (Project 31170) and by contributions from the provinces of British Columbia, Québec, and Ontario. The CHIME/FRB Project, which enabled development in common with the CHIME/Pulsar instrument, is funded by a grant from the CFI 2015 Innovation Fund (Project 33213) and by contributions from the provinces of British Columbia and Québec, and by the Dunlap Institute for Astronomy and Astrophysics at the University of Toronto. Additional support was provided by the Canadian Institute for Advanced Research (CIFAR), McGill University, and the McGill Space Institute thanks to the Trottier Family Foundation and the University of British Columbia. The CHIME/Pulsar instrument hardware was funded by NSERC RTI-1 grant No. EQPEQ 458893-2014.

J.K.S., D.L.K., M.A.M., M.E.D., T.D., S.M.R., and X.S. are supported by the NANOGrav NSF Physics Frontiers Center award numbers 1430284 and 2020265. E.P. is supported by an H2020 ERC Consolidator Grant 'MAGNESIA' under grant agreement No. 817661 and National Spanish grant PGC2018-095512-BI00. Z.P. is a Dunlap Fellow. J.vL. acknowledges funding from the European Research Council under the European Unions Seventh Framework Programme (FP/2007-2013)/ERC grant agreement No. 617199 ("ALERT"), and from Vici research program "ARGO" with project number 639.043.815, financed by the Netherlands Organisation for Scientific Research (NWO). M.A.M. and E.F.L. are supported by NSF OIA-1458952 and NSF Award Number 2009425. S.M. R. is a CIFAR Fellow. Pulsar research at UBC is supported by an NSERC Discovery Grant and by the Canadian Institute for Advanced Research. M.S. acknowledges funding from the European Research Council (ERC) under the European Unions Horizon 2020 research and innovation program (grant agreement No. 694745). M.E.D. acknowledges support from the Naval Research Laboratory by NASA under contract S-15633Y. T.D. is supported by an NSF Astronomy and Astrophysics Grant (AAG) award number 2009468. K.W.M. holds the Adam J. Burgasser Chair in Astrophysics and is supported by NSF grant Nos. 2008031 and 2018490. F.A.D is supported by a UBC Four Year Fellowship.

This research was enabled in part by support provided by the BC Digital Research Infrastructure Group and the Digital Research Alliance of Canada (https://alliancecan.ca).

### Appendix
### Timing Residuals

For each of the 128 pulsars timed as a part of this work, we include full timing residuals (units of milliseconds) in Figures 13–34. TOAs and error bars are colored by the observatory where they were measured: blue x symbols indicate CHIME, green plus sign markers indicate GBT, yellow arrows indicate Arecibo, and red circles indicate LOFAR.

---

[32] https://github.com/GBNCC/data

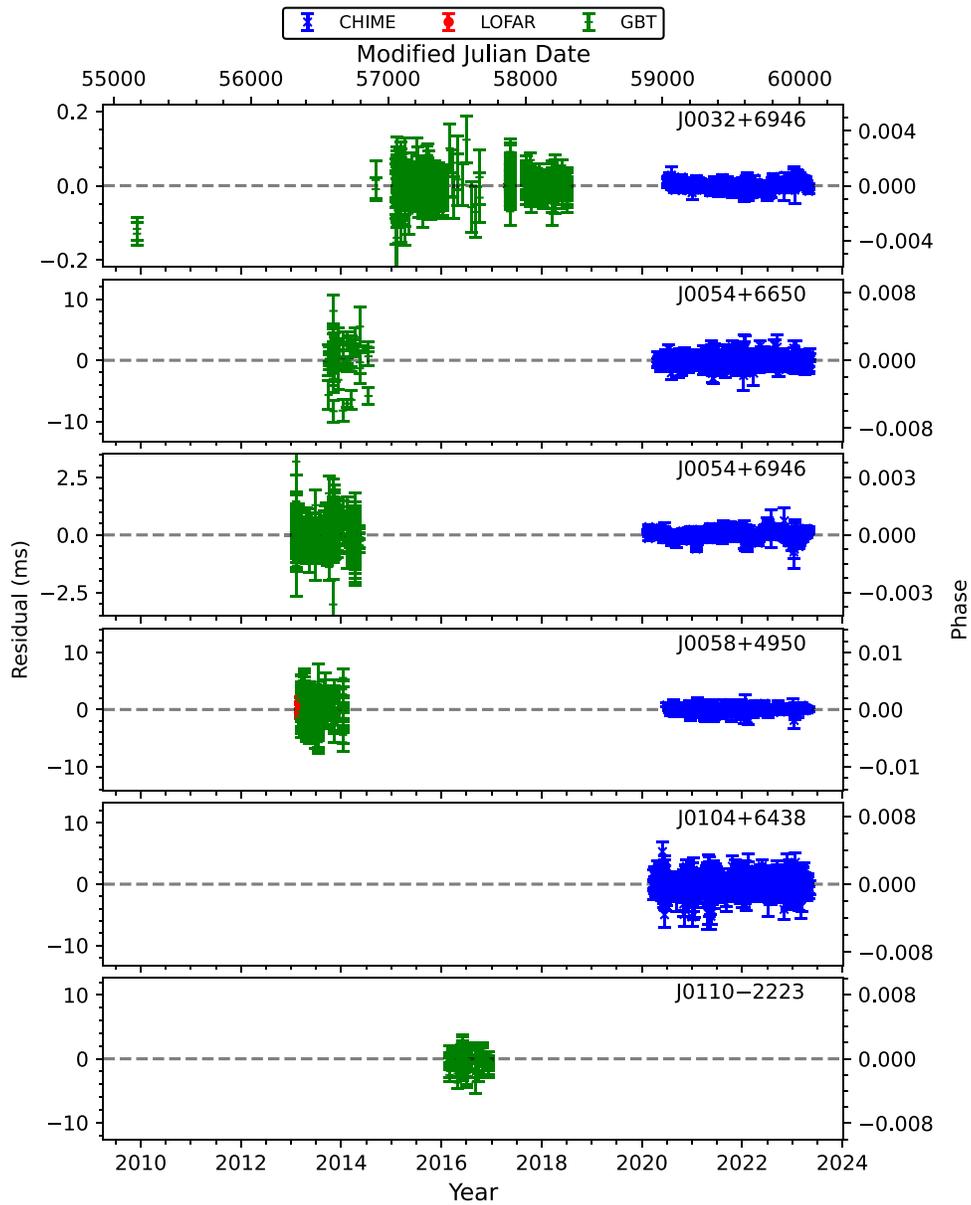

**Figure 13.** Timing residuals for sources in this study. TOAs are colored by their observatory: blue corresponds to CHIME, green to GBT, red to LOFAR, and yellow to Arecibo.

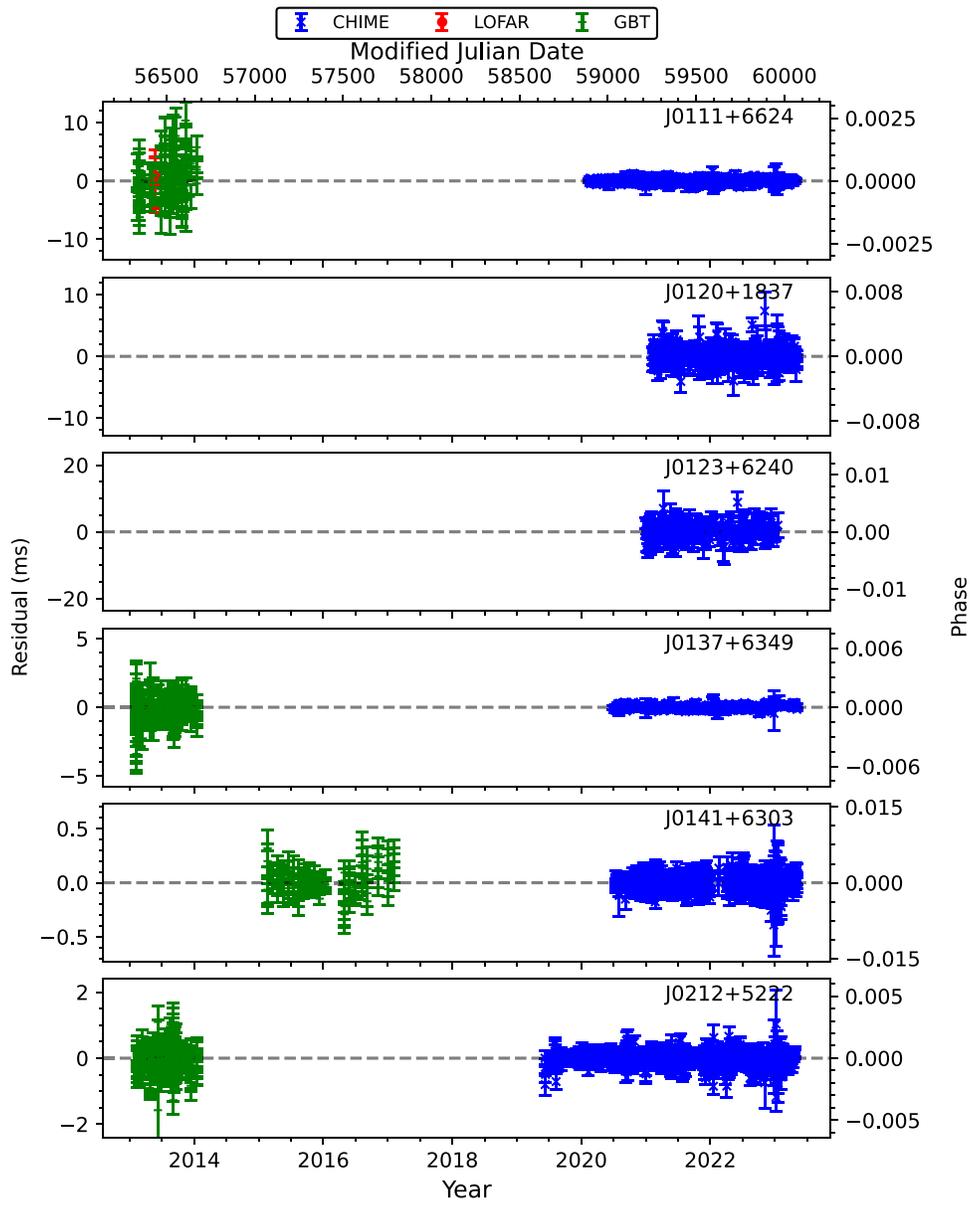

Figure 14. Timing residuals (continued). See Figure 13 for details.

  McEwen et al.

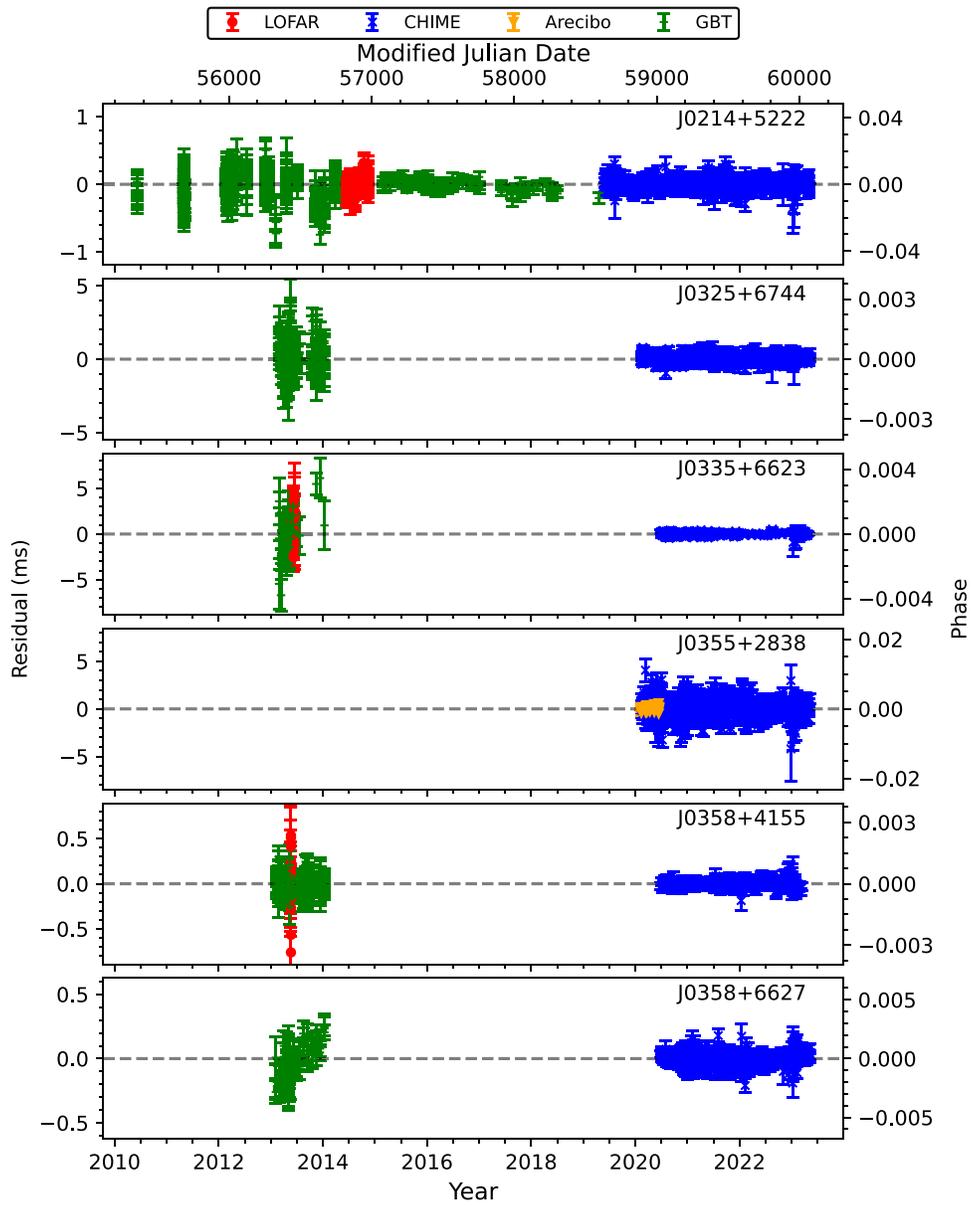

Figure 15. Timing residuals (continued). See Figure 13 for details.

    McEwen et al.

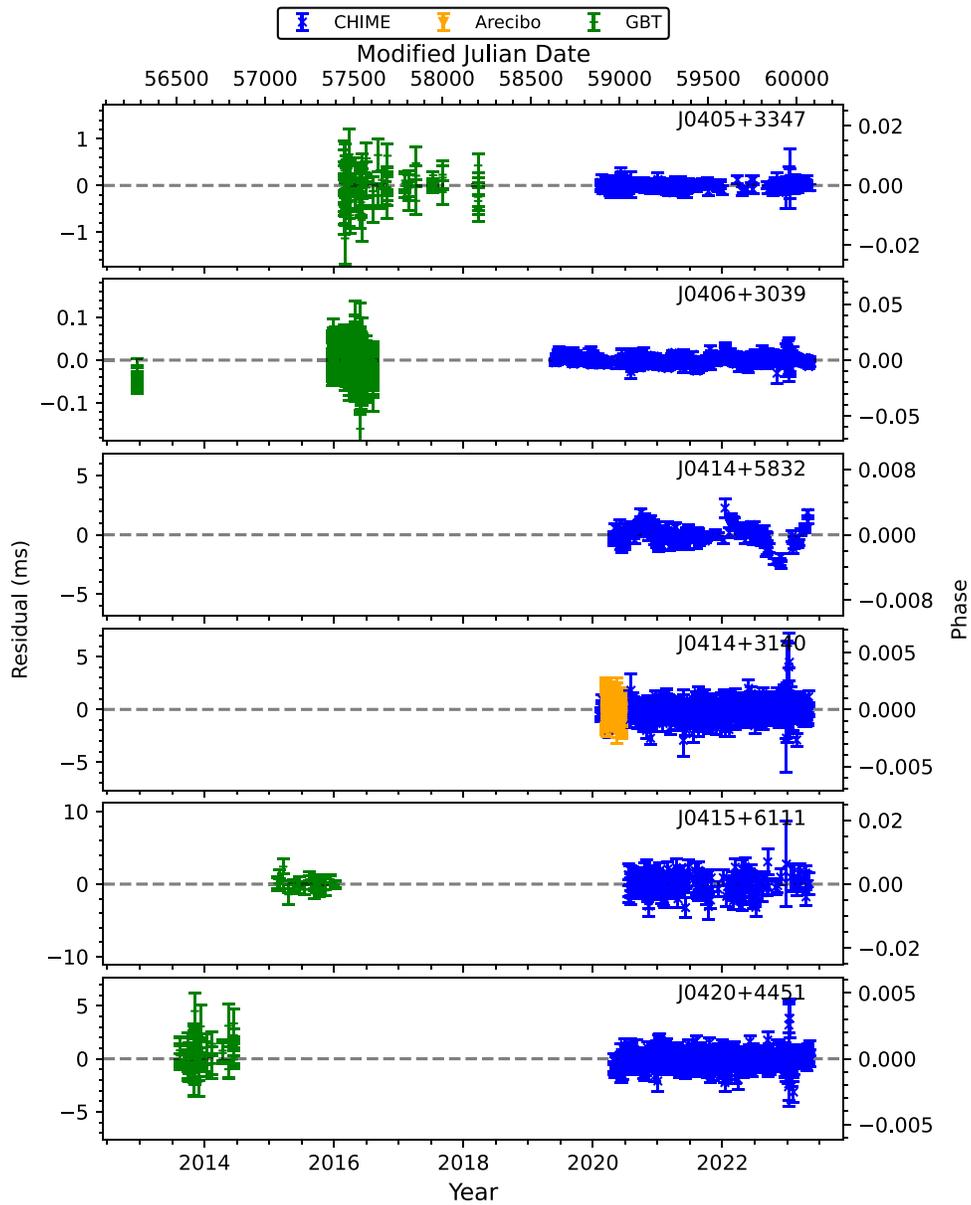

**Figure 16.** Timing residuals (continued). See Figure 13 for details.

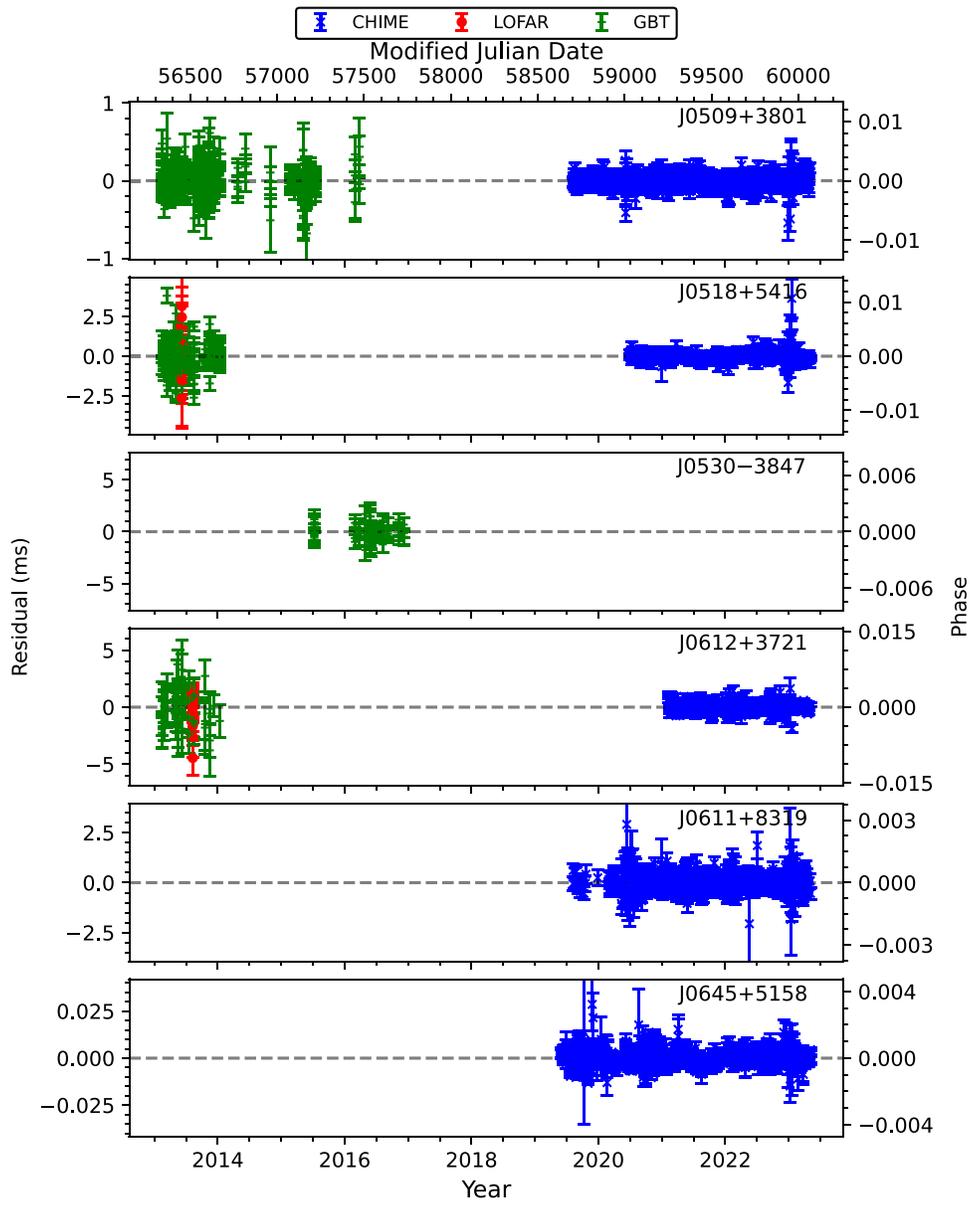

**Figure 17.** Timing residuals (continued). See Figure 13 for details.

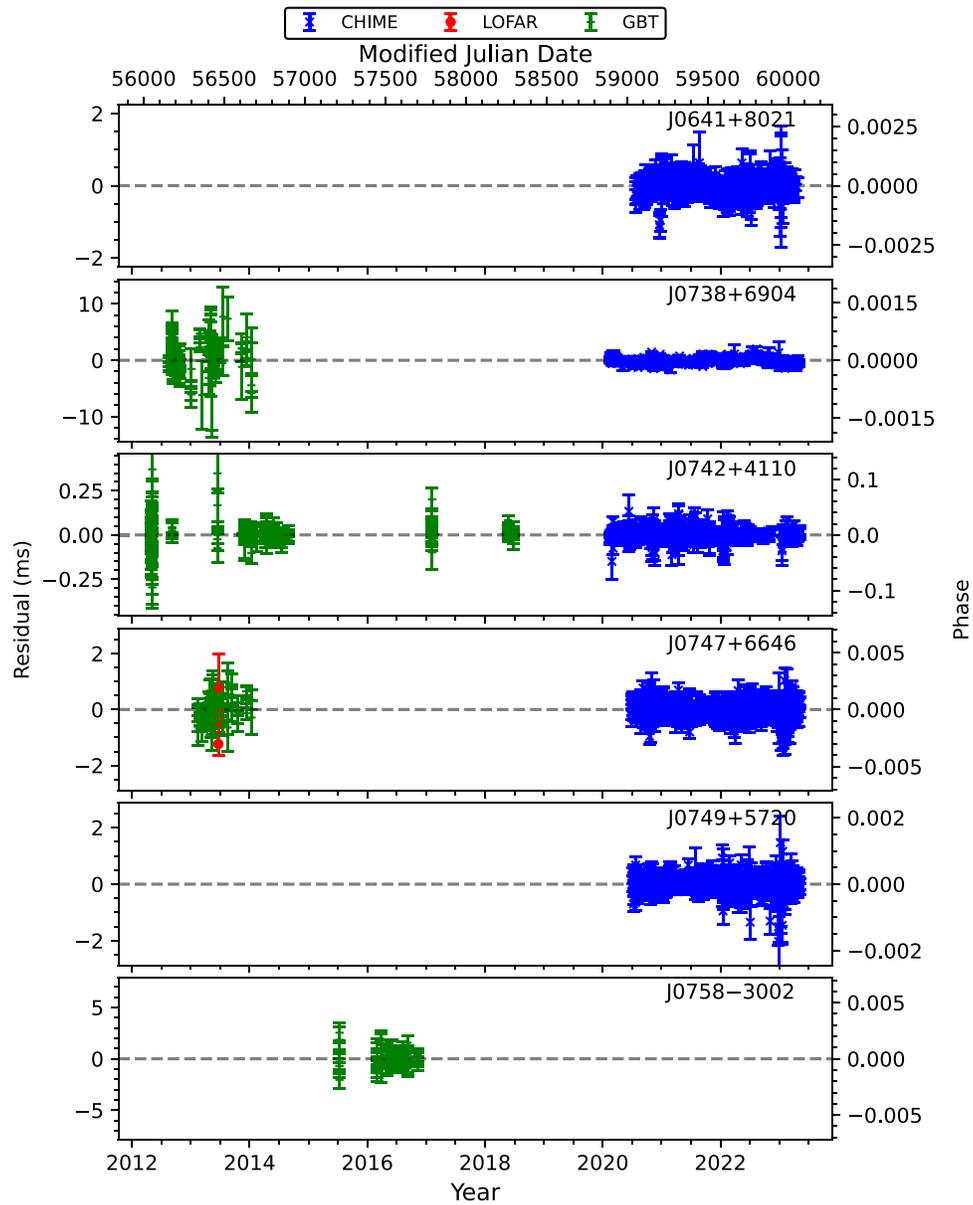

Figure 18. Timing residuals (continued). See Figure 13 for details.

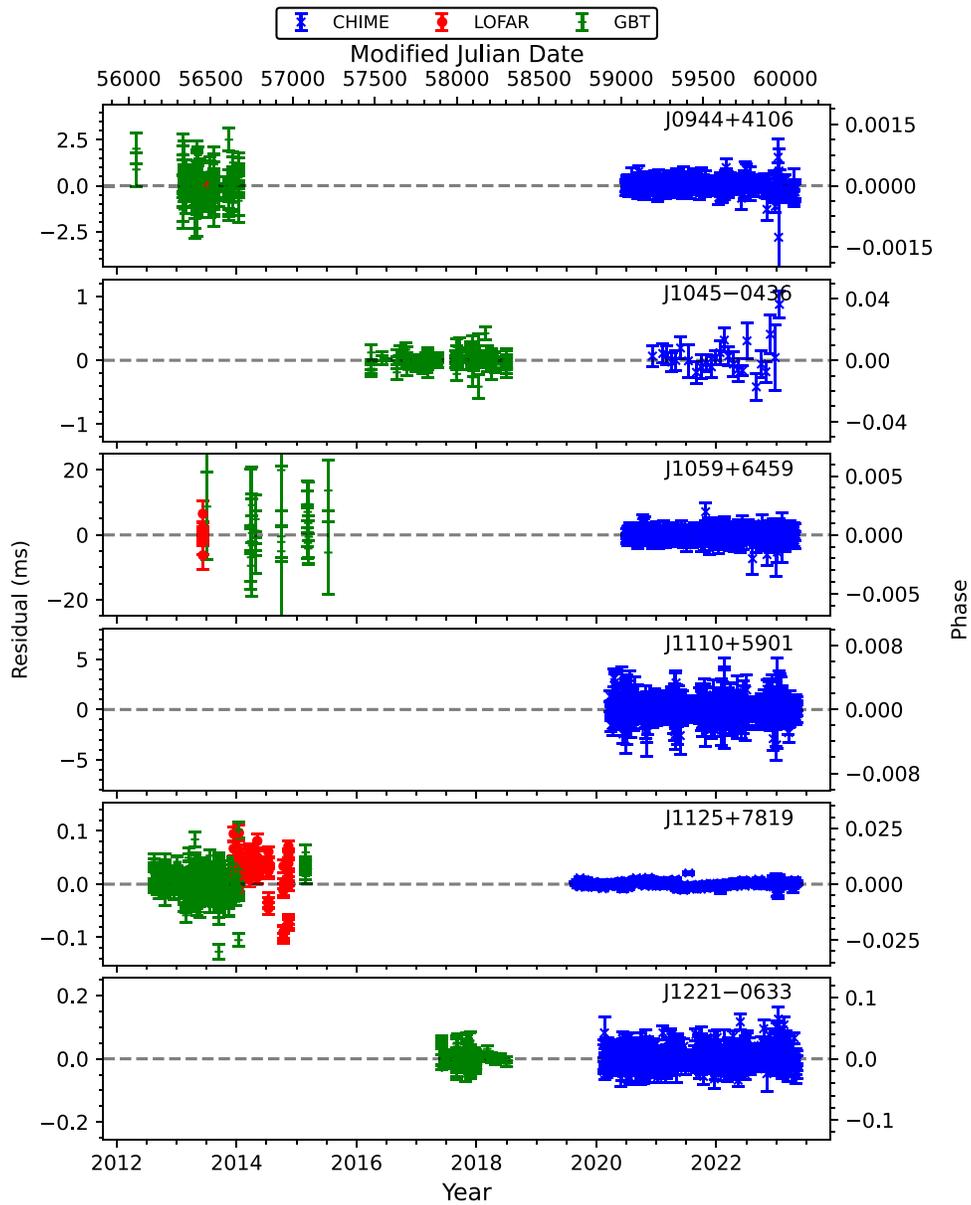

**Figure 19.** Timing residuals (continued). See Figure 13 for details.

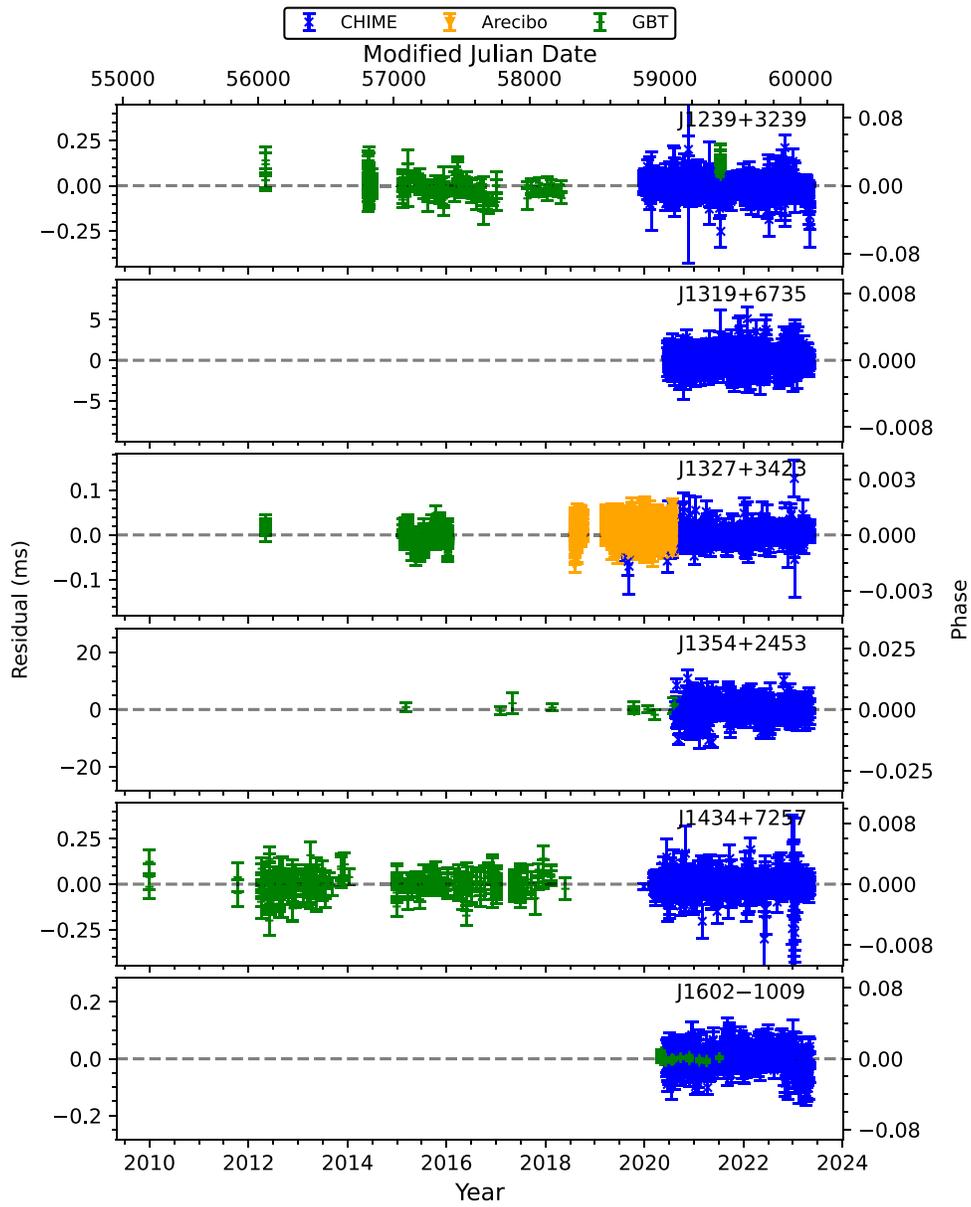

**Figure 20.** Timing residuals (continued). See Figure 13 for details.

 

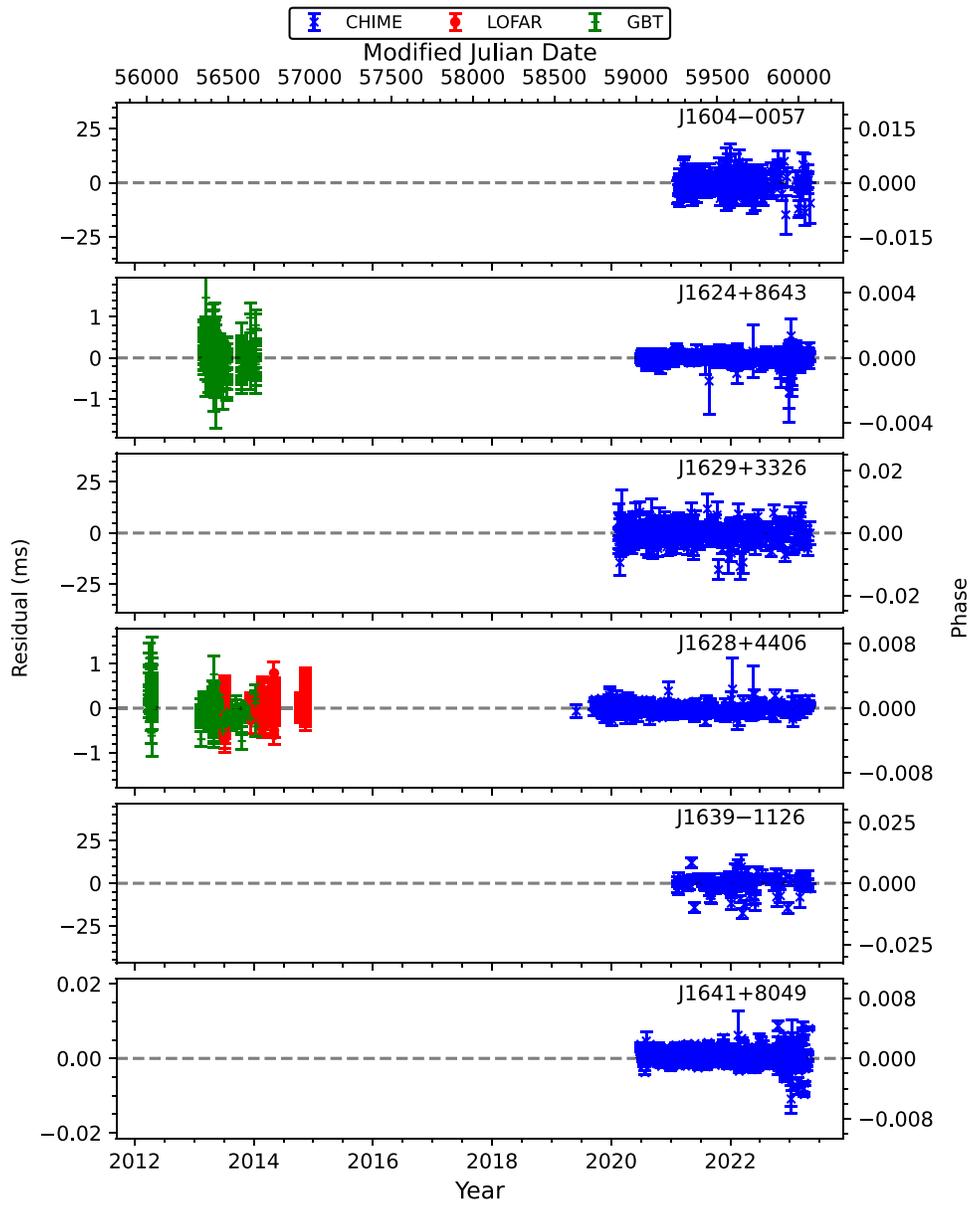

**Figure 21.** Timing residuals (continued). See Figure 13 for details.

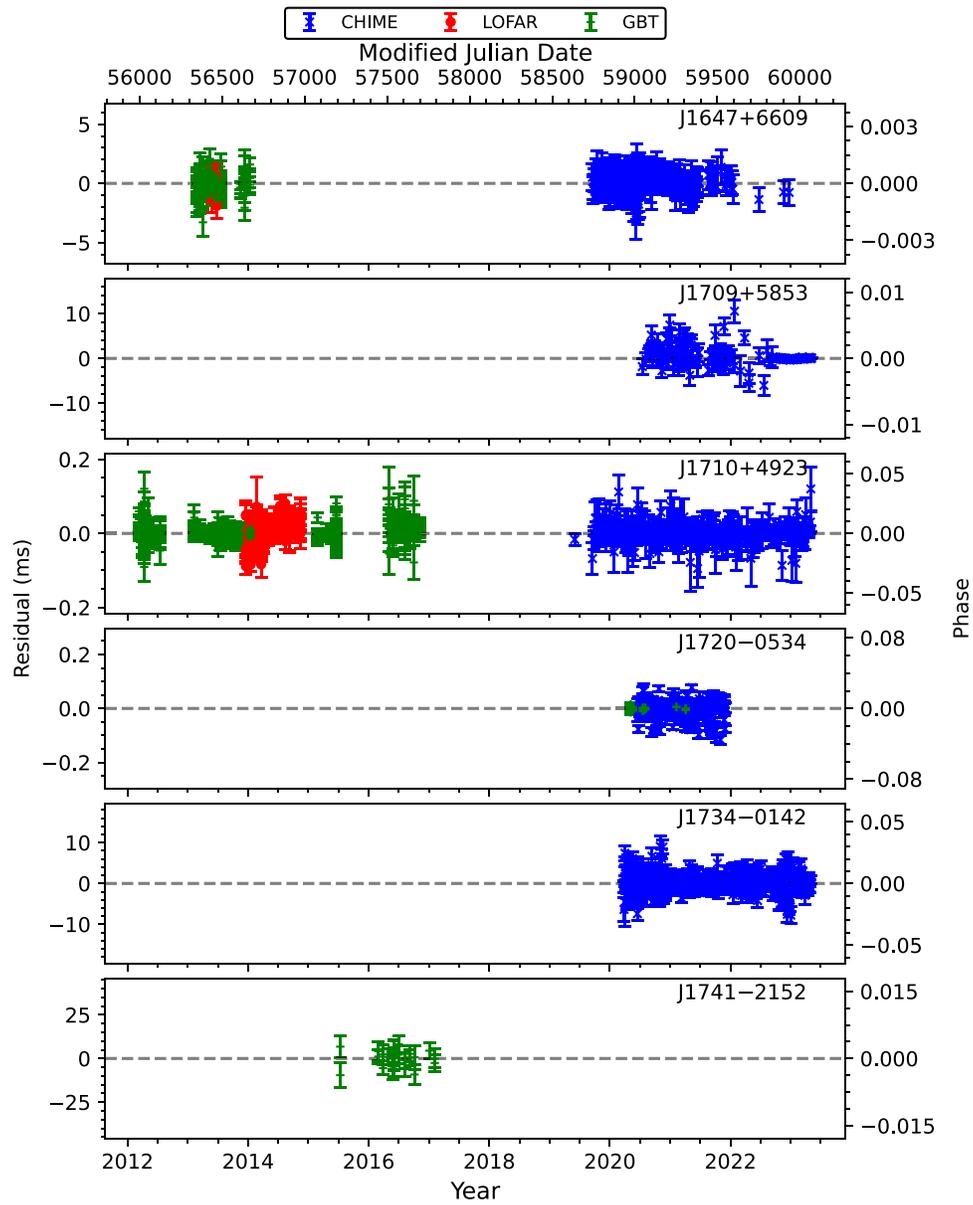

**Figure 22.** Timing residuals (continued). See Figure 13 for details. For PSR J1709+5853, an improved position determined from timing dramatically improved TOA precision near the beginning of 2022.

 

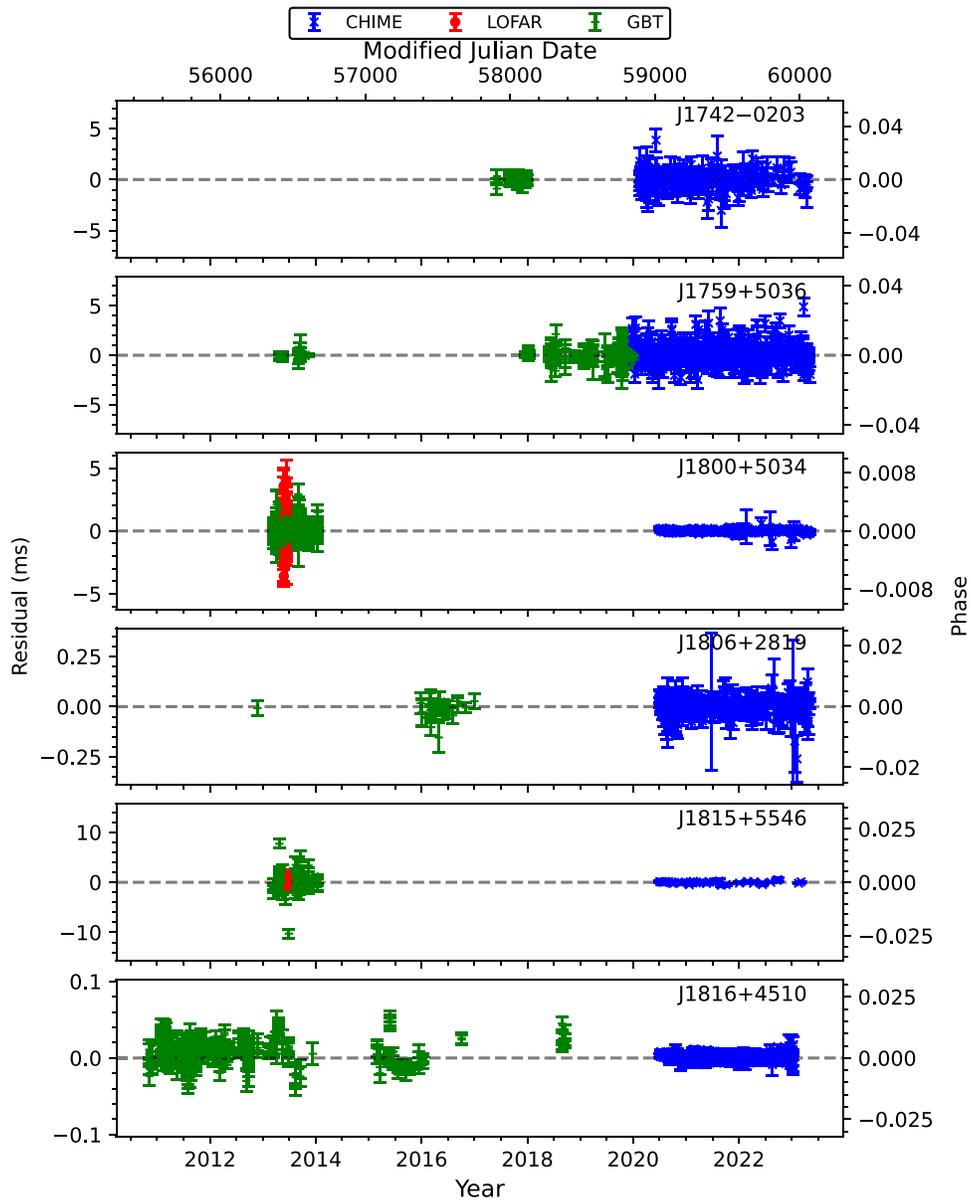

**Figure 23.** Timing residuals (continued). See Figure 13 for details.

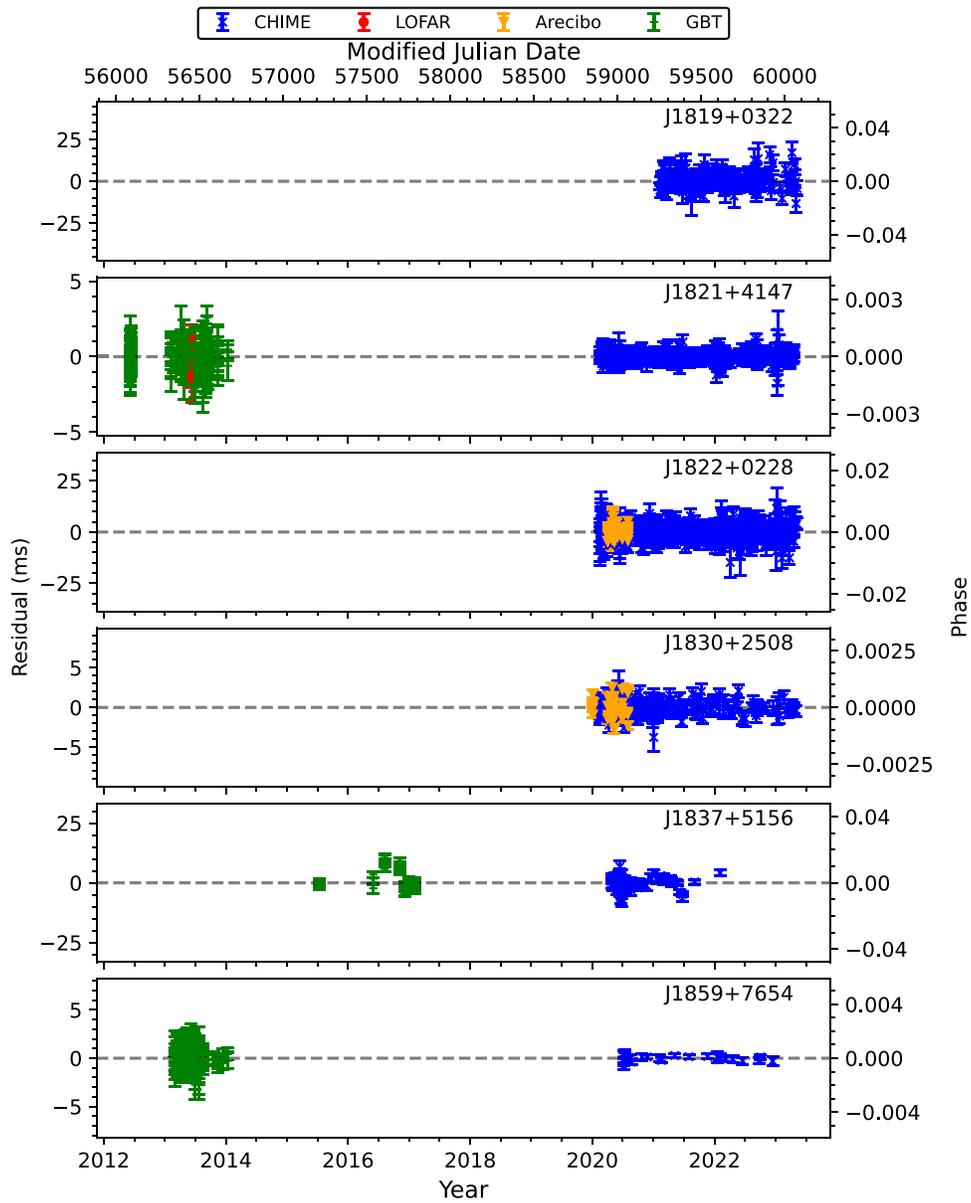

**Figure 24.** Timing residuals (continued). See Figure 13 for details.

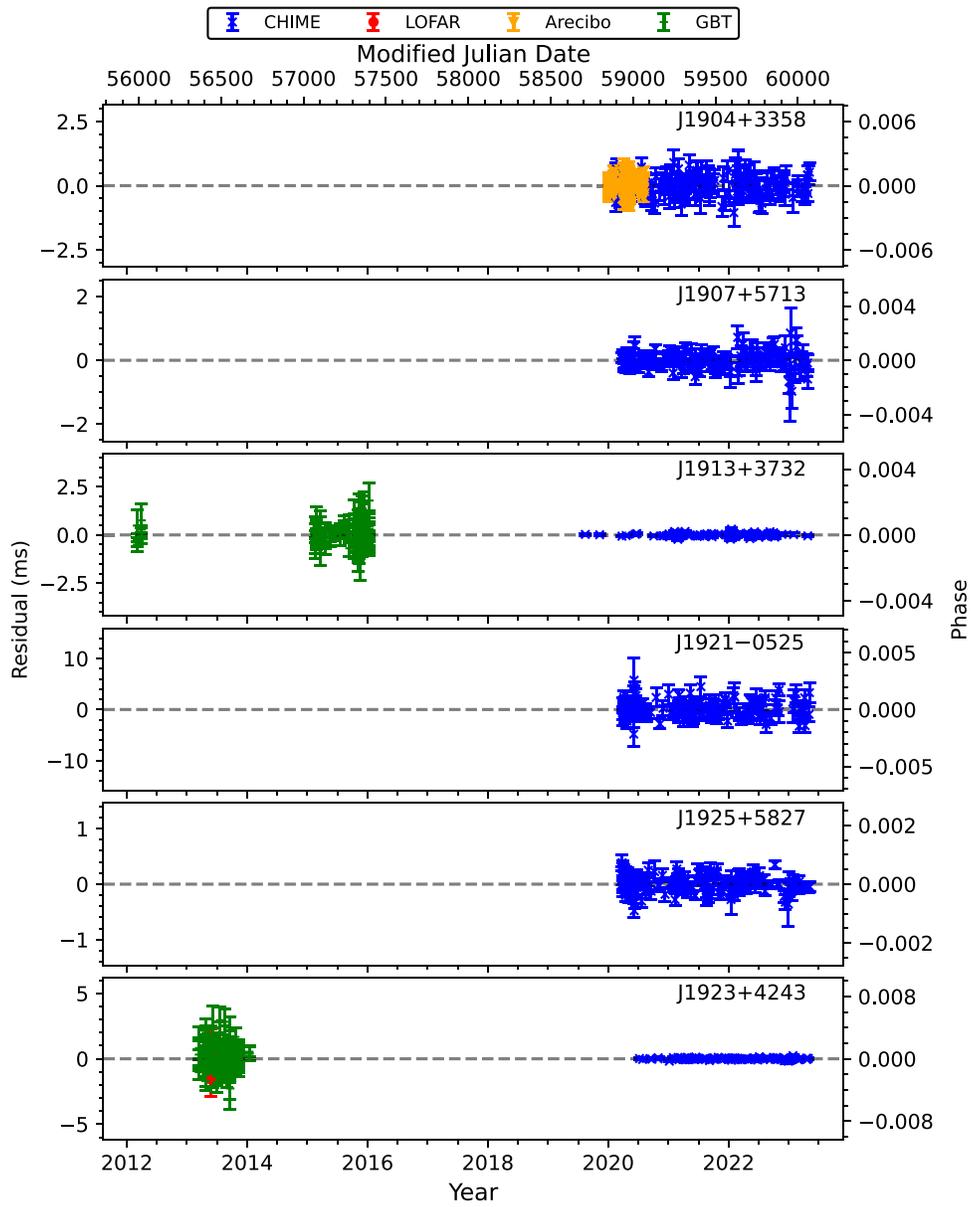

**Figure 25.** Timing residuals (continued). See Figure 13 for details.

The Astrophysical Journal, 962:167 (39pp), 2024 February 20McEwen et al.

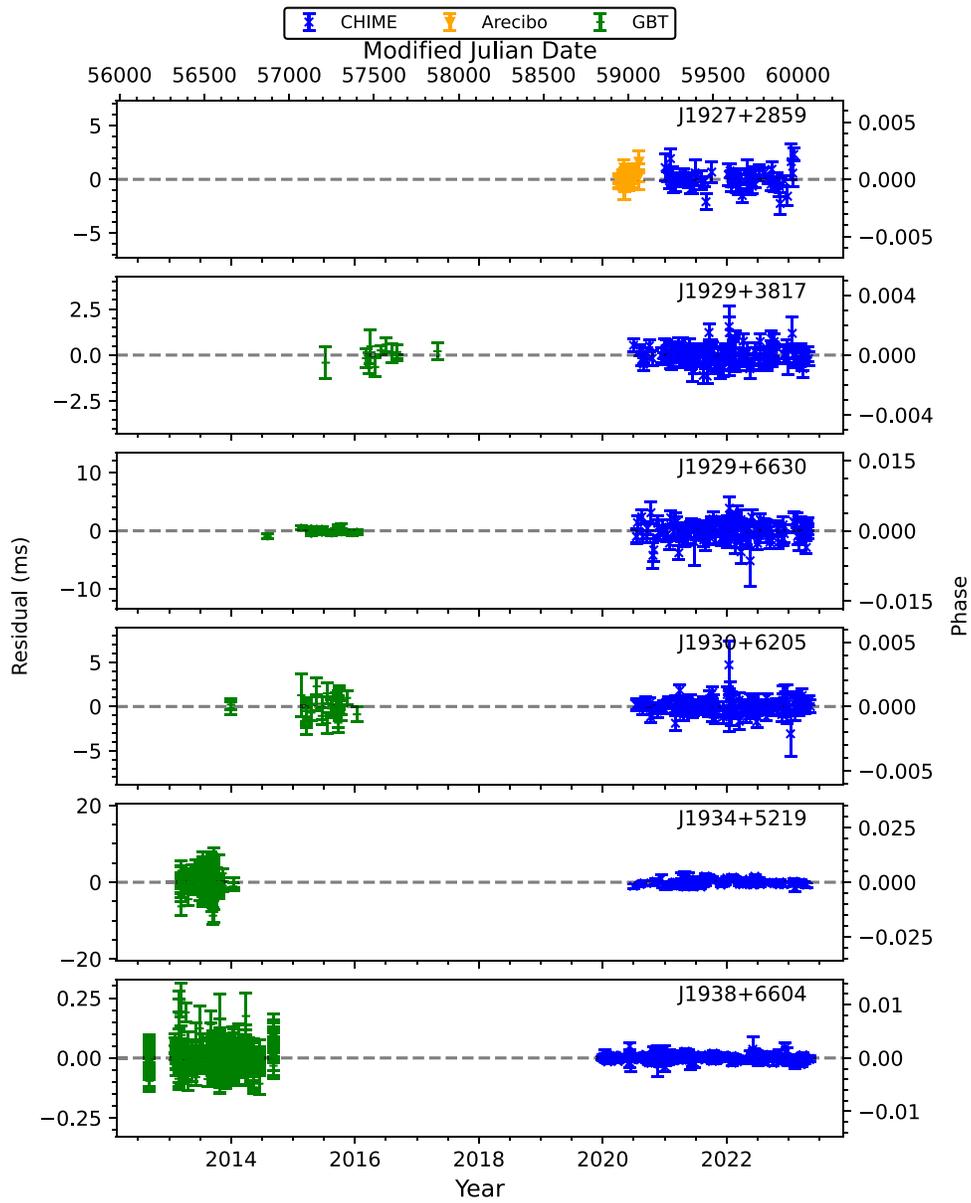

**Figure 26.** Timing residuals (continued). See Figure 13 for details.

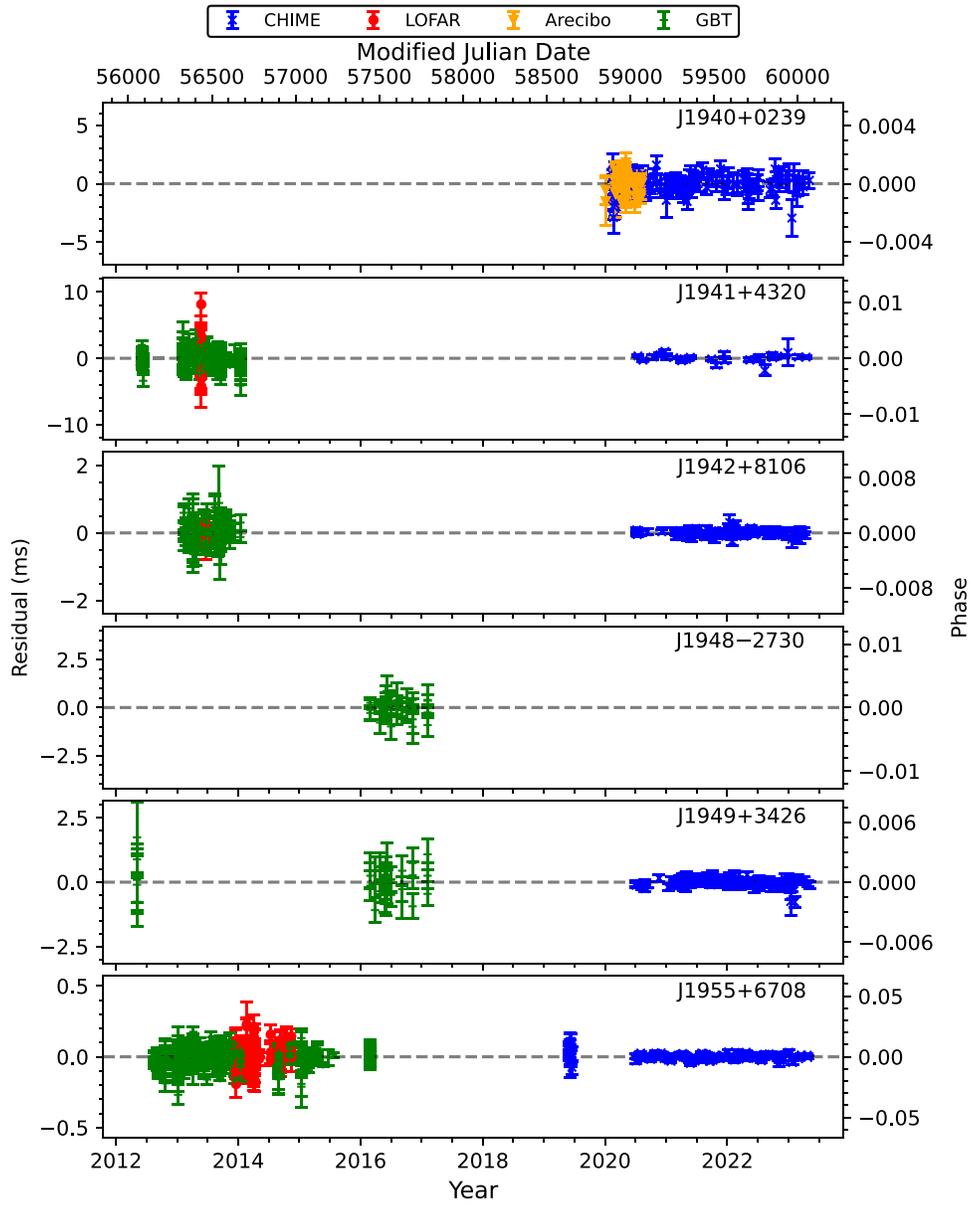

**Figure 27.** Timing residuals (continued). See Figure 13 for details.

 McEwen et al.

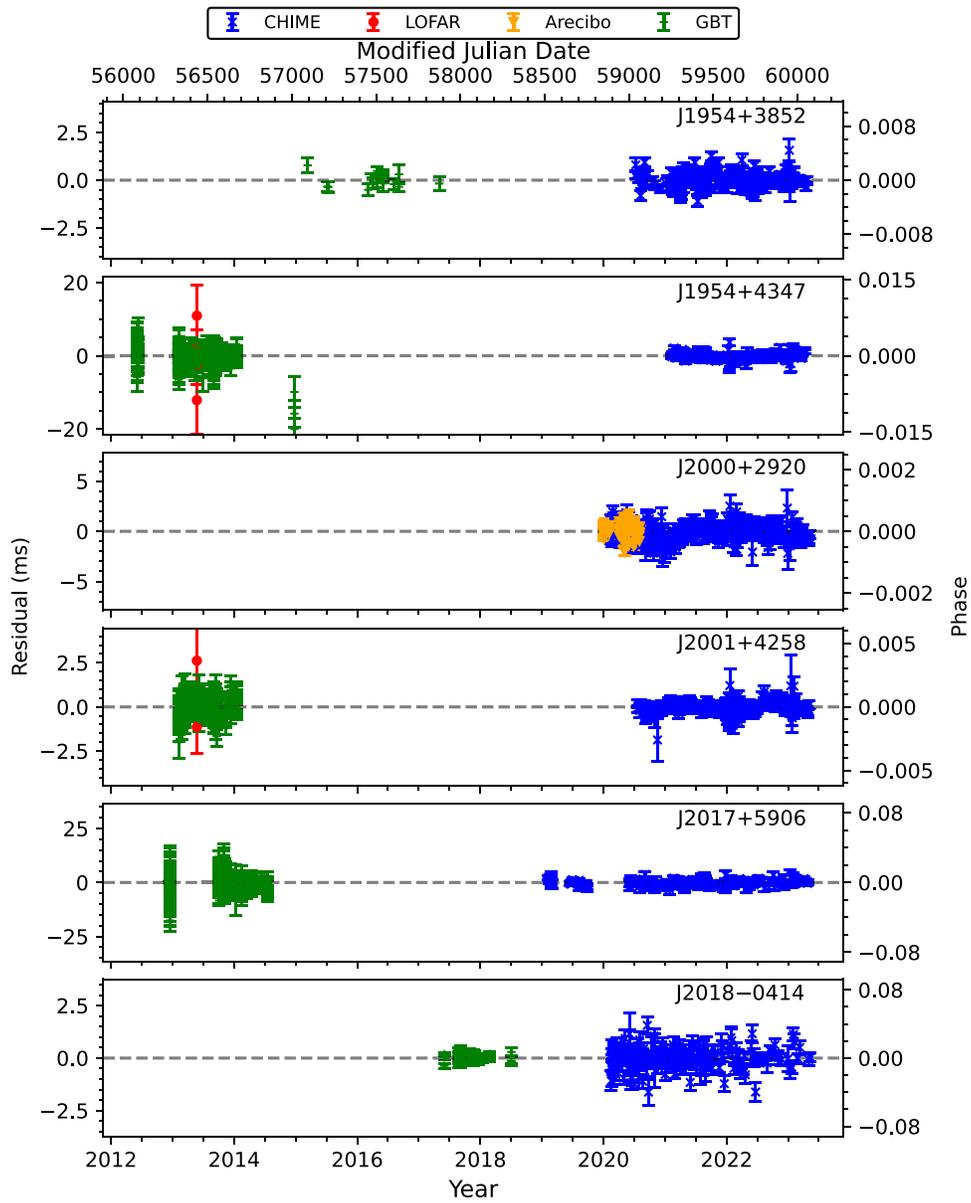

**Figure 28.** Timing residuals (continued). See Figure 13 for details.

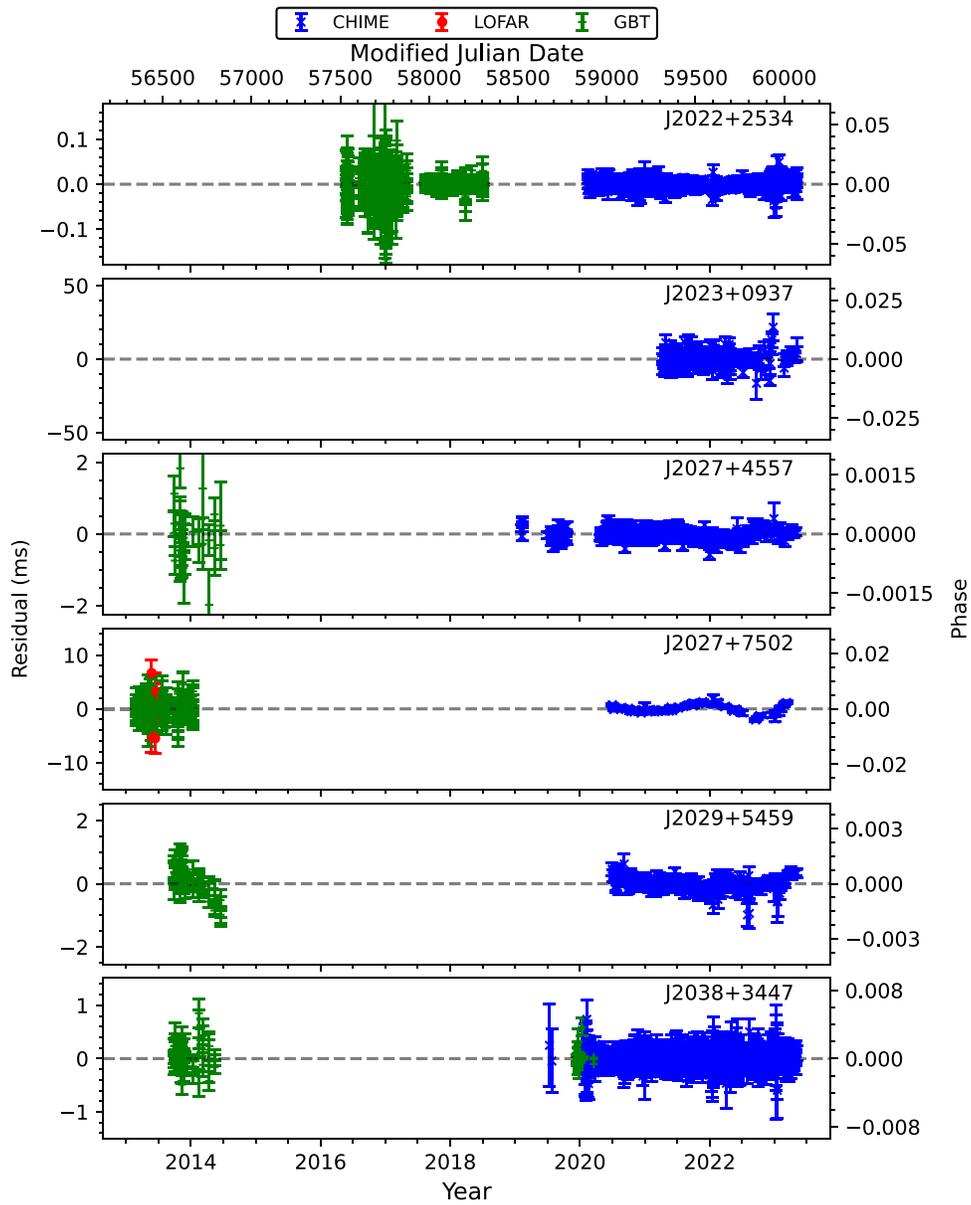

Figure 29. Timing residuals (continued). See Figure 13 for details.

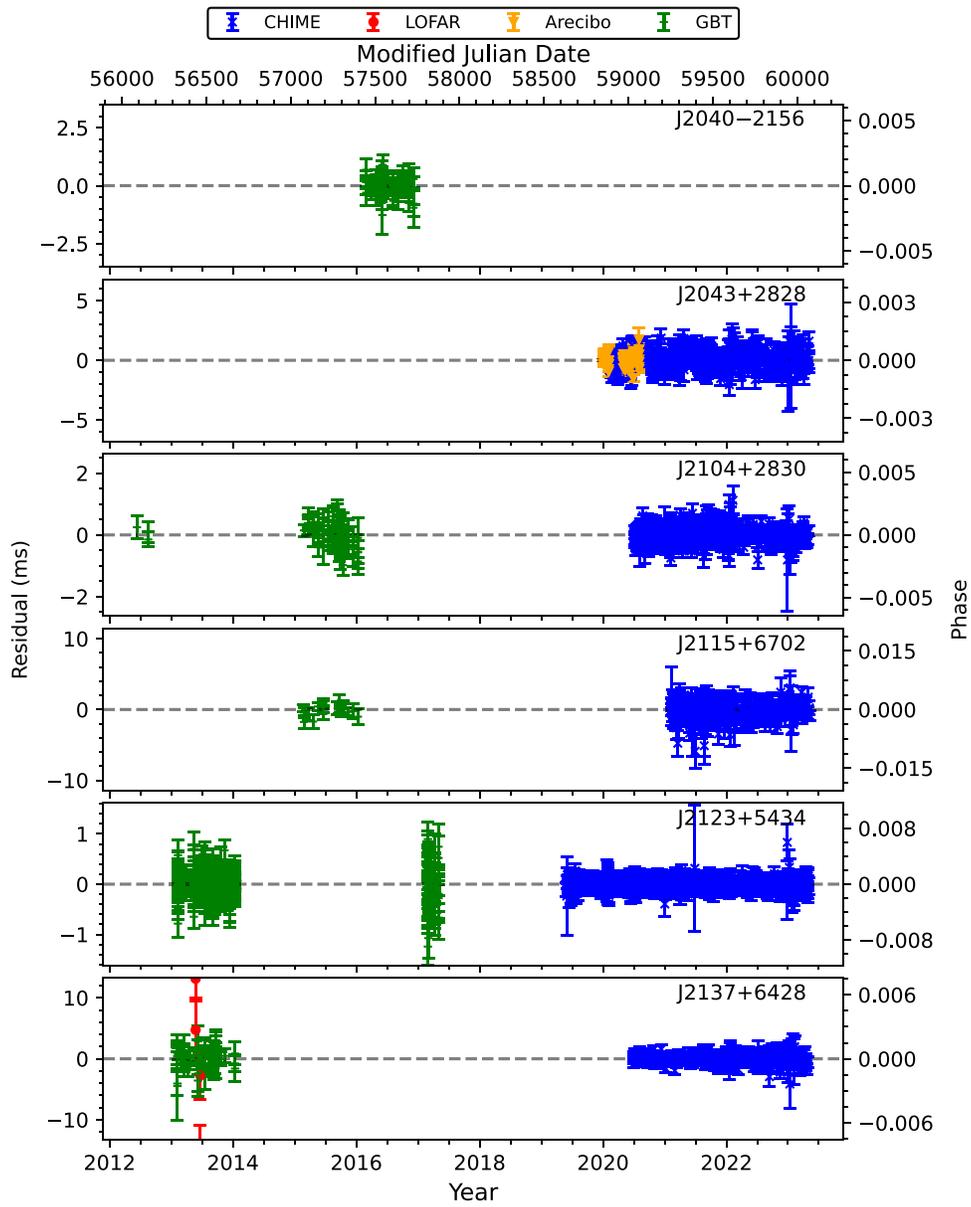

Figure 30. Timing residuals (continued). See Figure 13 for details.

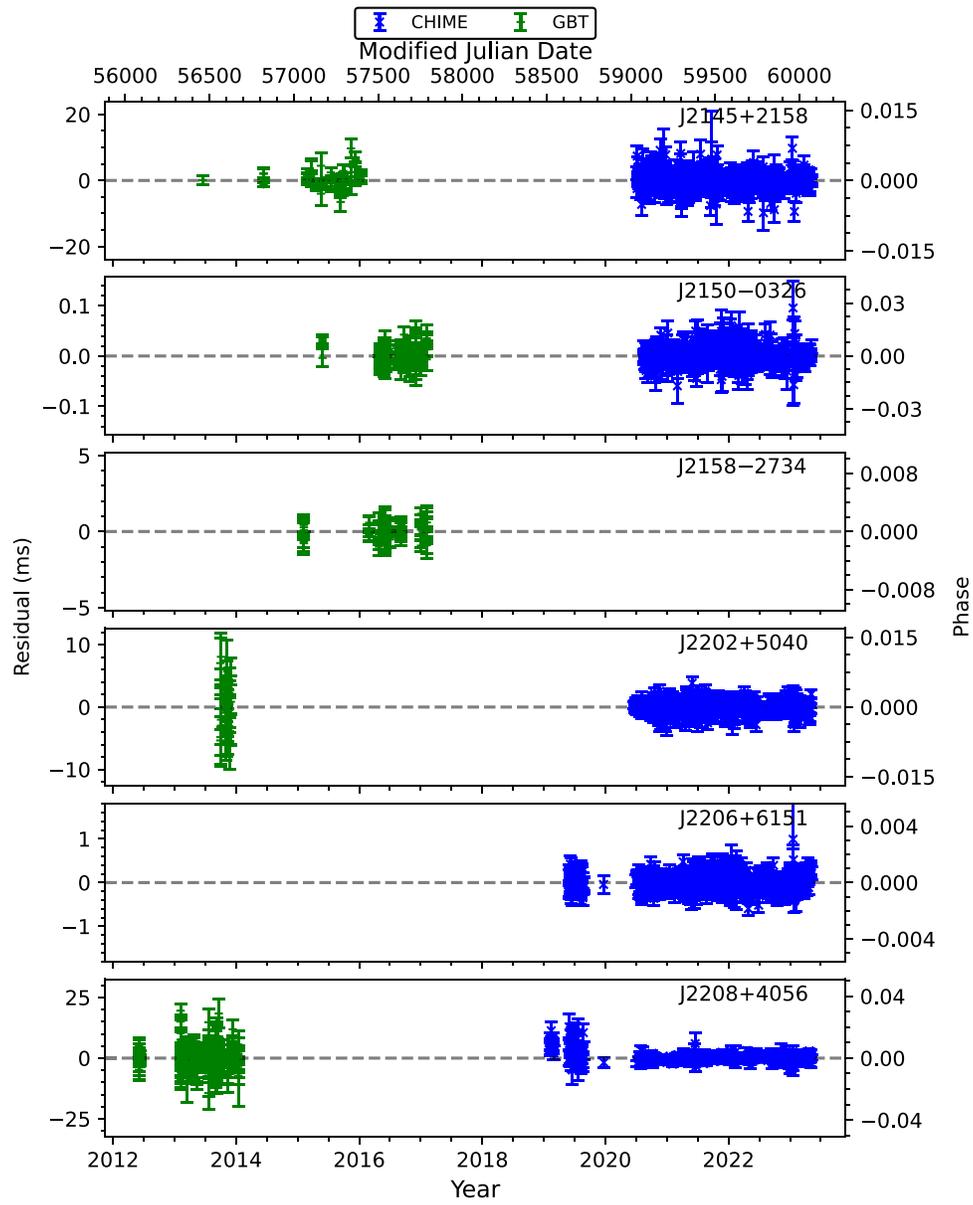

Figure 31. Timing residuals (continued). See Figure 13 for details.

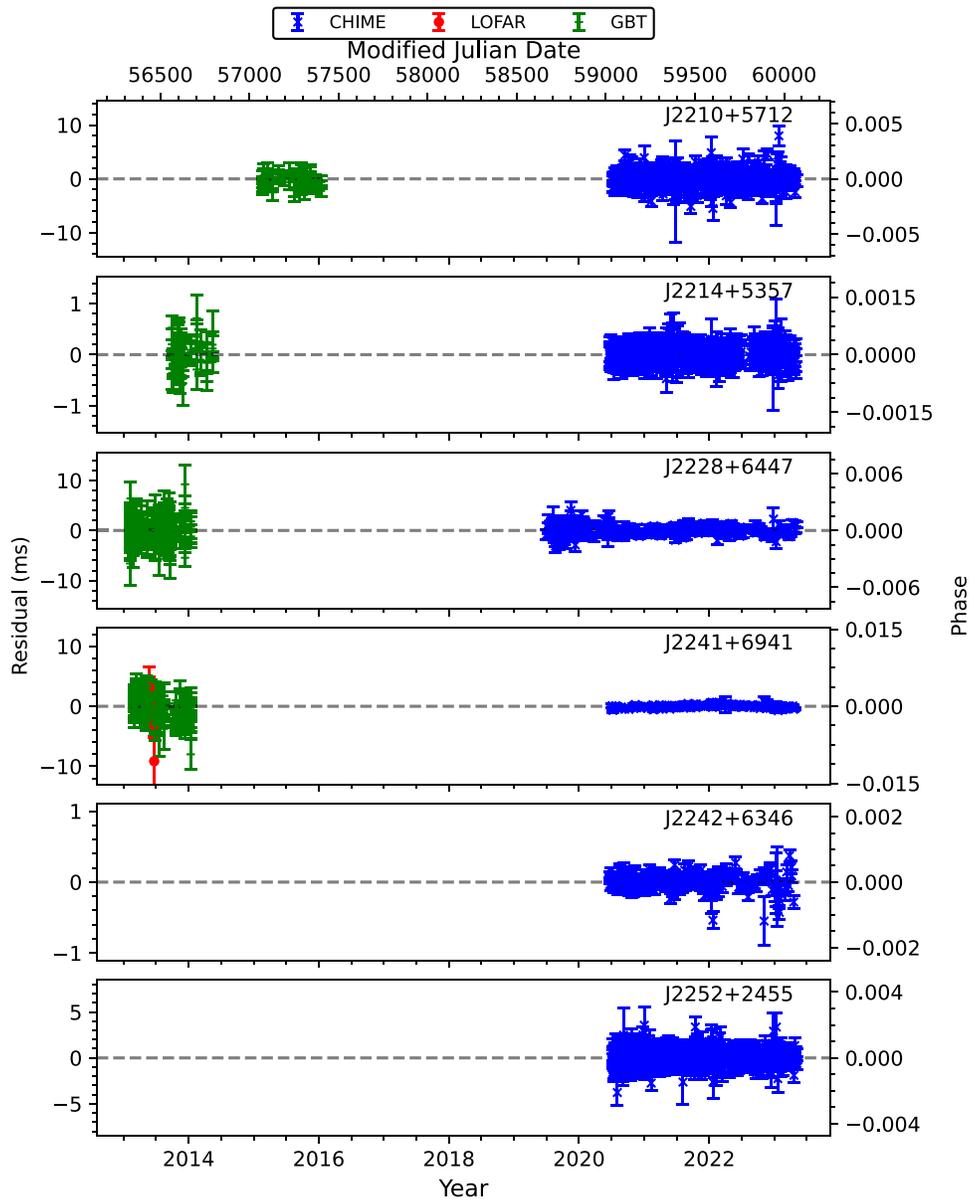

Figure 32. Timing residuals (continued). See Figure 13 for details.

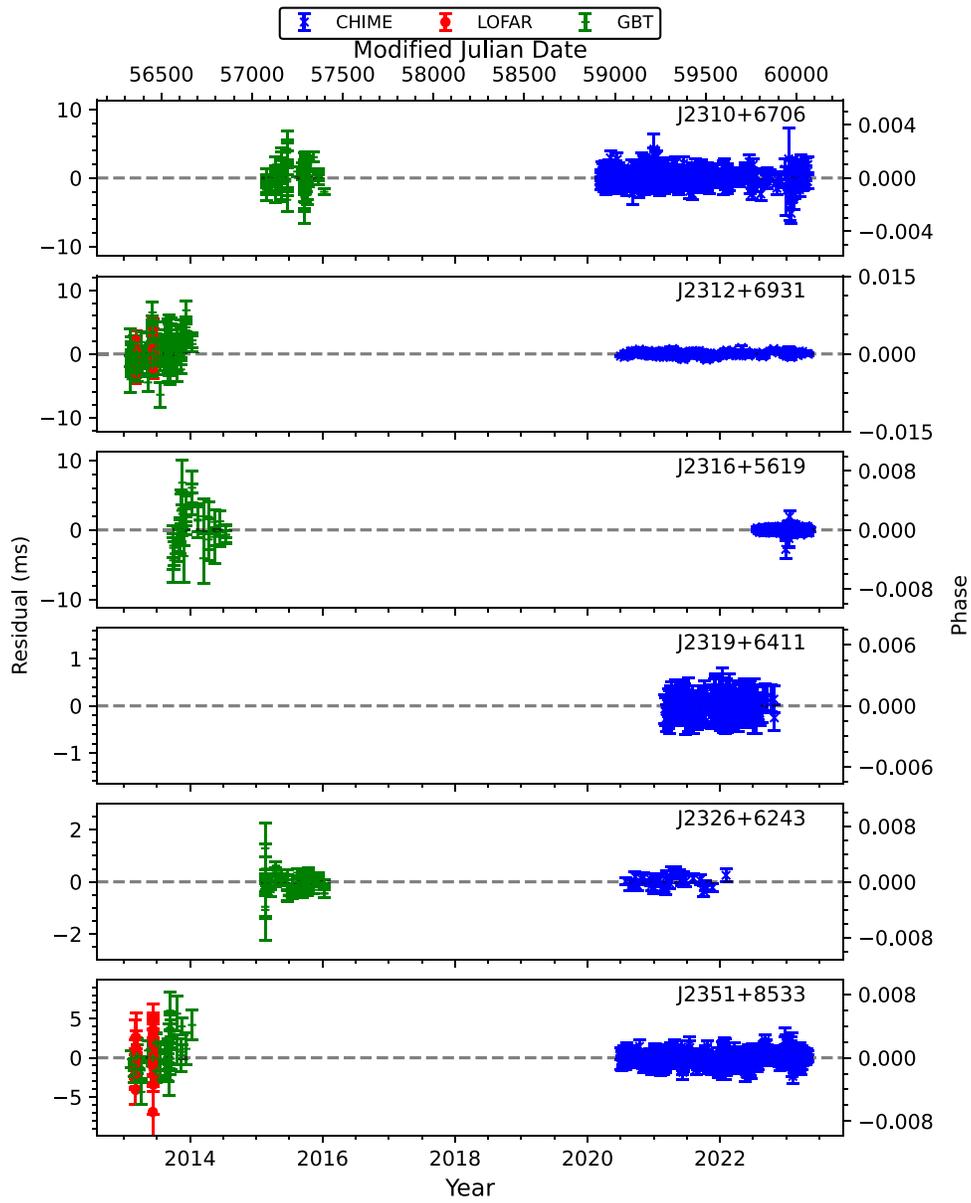

**Figure 33.** Timing residuals (continued). See Figure 13 for details.

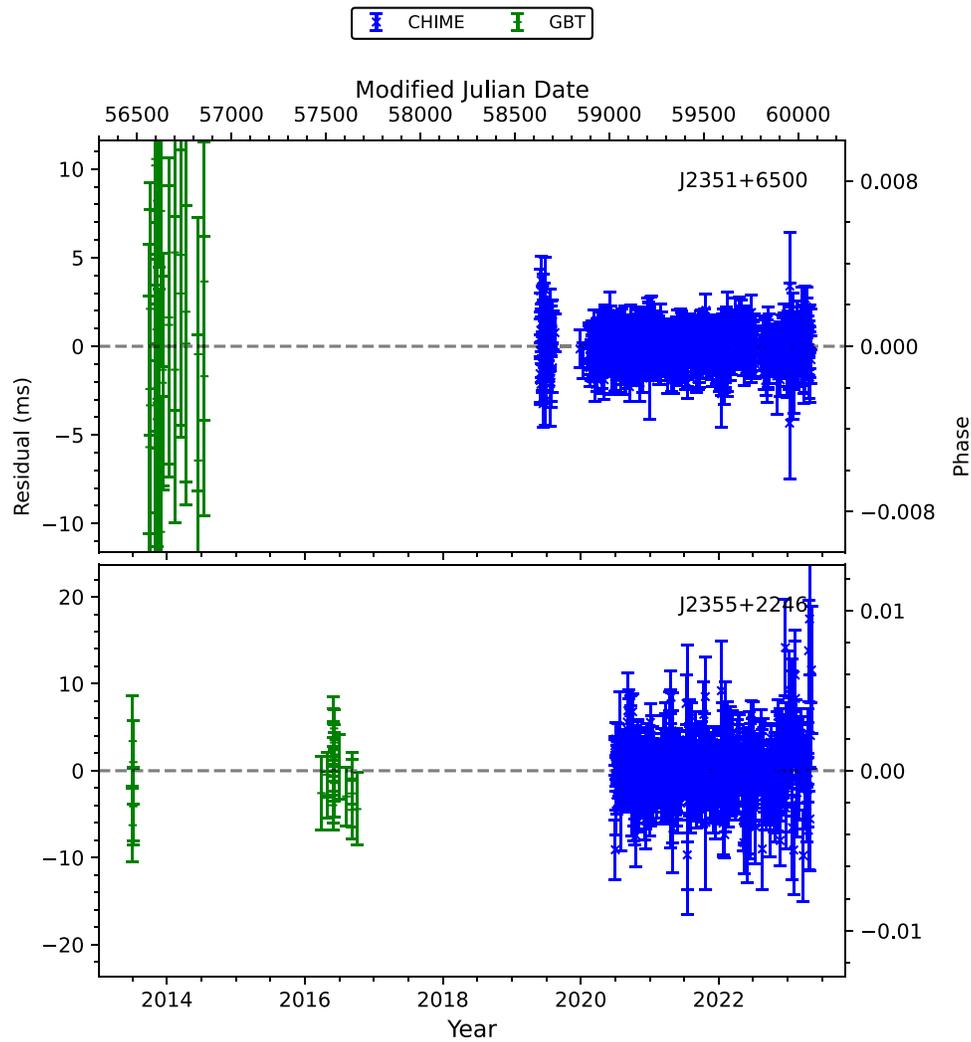

**Figure 34.** Timing residuals (continued). See Figure 13 for details.

**ORCID iDs**

A. E. McEwen ⓘ https://orcid.org/0000-0001-5481-7559
J. K. Swiggum ⓘ https://orcid.org/0000-0002-1075-3837
D. L. Kaplan ⓘ https://orcid.org/0000-0001-6295-2881
C. M. Tan ⓘ https://orcid.org/0000-0001-7509-0117
B. W. Meyers ⓘ https://orcid.org/0000-0001-8845-1225
E. Fonseca ⓘ https://orcid.org/0000-0001-8384-5049
G. Y. Agazie ⓘ https://orcid.org/0000-0001-5134-3925
P. Chawla ⓘ https://orcid.org/0000-0002-3426-7606
K. Crowter ⓘ https://orcid.org/0000-0002-1529-5169
M. E. DeCesar ⓘ https://orcid.org/0000-0002-2185-1790
T. Dolch ⓘ https://orcid.org/0000-0001-8885-6388
F. A. Dong ⓘ https://orcid.org/0000-0003-4098-5222
W. Fiore ⓘ https://orcid.org/0000-0001-5645-5336
D. C. Good ⓘ https://orcid.org/0000-0003-1884-348X
A. G. Istrate ⓘ https://orcid.org/0000-0002-8811-8171
V. M. Kaspi ⓘ https://orcid.org/0000-0001-9345-0307
V. I. Kondratiev ⓘ https://orcid.org/0000-0001-8864-7471
J. van Leeuwen ⓘ https://orcid.org/0000-0001-8503-6958
L. Levin ⓘ https://orcid.org/0000-0002-2034-2986
E. F. Lewis ⓘ https://orcid.org/0000-0002-2972-522X
R. S. Lynch ⓘ https://orcid.org/0000-0001-5229-7430
K. W. Masui ⓘ https://orcid.org/0000-0002-4279-6946
J. W. McKee ⓘ https://orcid.org/0000-0002-2885-8485
M. A. McLaughlin ⓘ https://orcid.org/0000-0001-7697-7422
H. Al Noori ⓘ https://orcid.org/0000-0002-4187-4981
E. Parent ⓘ https://orcid.org/0000-0002-0430-6504
S. M. Ransom ⓘ https://orcid.org/0000-0001-5799-9714
X. Siemens ⓘ https://orcid.org/0000-0002-7778-2990
R. Spiewak ⓘ https://orcid.org/0000-0002-6730-3298
I. H. Stairs ⓘ https://orcid.org/0000-0001-9784-8670